\documentclass[11pt]{article}
\usepackage[UKenglish]{babel}
\usepackage[a4paper, margin=1in]{geometry}

\usepackage{hyperref}
\usepackage{graphicx}
\usepackage{caption}
\usepackage{subcaption}
\usepackage{amsmath}
\usepackage{amssymb}
\usepackage{mathtools}
\usepackage{siunitx}
\usepackage{multirow}
\usepackage{xcolor}
\usepackage{ctable}
\usepackage{bm}
\usepackage{upgreek}

\graphicspath{{figures/}}

\title{Exploring descriptors for titanium microstructure via digital fingerprints from variational autoencoders}
\author{Michael D. White$^{1,}$\footnote{Corresponding author \newline \textit{Email address:} michael.white-3@postgrad.manchester.ac.uk (M.D. White)}, Gowtham Nimmal Haribabu$^{1, 3}$, Jeyapriya Thimukonda\\ Jegadeesan$^{3}$, Bikramjit Basu$^{3}$, Philip J. Withers$^{1, 2}$ and Chris P. Race$^{2, 4}$ \\ \\
\small $^{1}$\textit{Department of Materials, University of Manchester, Manchester, UK, M13 9PL} \\
\small $^{2}$\textit{Henry Royce Institute, University of Manchester, Manchester, UK, M13 9PL} \\
\small $^{3}$\textit{Materials Research Centre, Indian Institute of Science, Bangalore, India, 560012} \\
\small $^{4}$\textit{Department of Materials Science and Engineering, University of Sheffield, Sheffield, UK, S1 3JD} \\
}
\date{\small \today}

\begin{document}

\maketitle

\noindent\rule{\textwidth}{.5pt}
\begin{abstract}
    Microstructure is key to controlling and understanding the properties of metallic materials, but traditional approaches to describing microstructure capture only a small number of features.
    To enable data-centric approaches to materials discovery, to allow efficient storage of microstructural data and to assist in quality control in metals processing, we require more complete descriptors of microstructure.
    The concept of \emph{microstructural fingerprinting}, using machine learning (ML) to develop quantitative, low-dimensional descriptors of microstructures, has recently attracted significant attention.
    However, it is difficult to interpret conclusions drawn by ML algorithms, which are often referred to as ``black boxes''.
    For example, convolutional neural networks (CNNs) can be trained to make predictions about a material from a set of microstructural image data, but the feature space that is learned is often used uncritically and adopted without any validation.

    Here we explore the use of variational autoencoders (VAEs), comprising a pair of CNNs, which can be trained to produce microstructural fingerprints in a continuous latent space.
    The VAE architecture also permits the reconstruction of images from fingerprints, allowing us to explore how key features of microstructure are encoded in the latent space of fingerprints.
    We develop a VAE architecture based on ResNet18 and train it on two classes of Ti-6Al-4V optical micrographs (bimodal and lamellar) as an example of an industrially important alloy where microstructural control is critical to performance.
    The latent/feature space of fingerprints learned by the VAE is explored in several ways, including by supplying interpolated and randomly perturbed fingerprints to the trained decoder and via dimensionality reduction to explore the distribution and correlation of microstructural features within the latent space of fingerprints.

    We show that the fingerprints generated via the trained VAE exhibit smooth, interpolable behaviour with stability to local perturbations, supporting their suitability as general purpose descriptors for microstructure.
    We also show that key properties of the microstructures (volume fraction and grain size) are strongly correlated with position in the encoded feature space, supporting the use of VAE fingerprints for quantitative exploration of process-structure-property relationships.
\end{abstract}
\noindent\rule{\textwidth}{.5pt}

\newpage
\section{Introduction}
\label{sec:intro}

In the field of metallurgy and materials science, process-structure-property (PSP) linkages are instrumental in guiding material design for targeted applications~\cite{basu2022, holm2020}.
Despite the importance of PSP linkages, a rigorous mathematical framework is not currently available for systematic analysis in this context~\cite{withers2019}.
The key problem is that, whilst compositional processing parameters and measured properties are inherently described by numbers, microstructure lacks a comprehensive numerical descriptor.
A central impediment is that the characteristic microstructural features exhibit heterogeneity over a wide range of size, spatial, and temporal scales.
From the application standpoint, it is important to identify a subset of salient measures of internal structure that can be tracked through a material's processing history, that capture the dominant influences on the targeted properties.
In conventional microstructural analysis, some quantitative methods are used, but these generally rely on metrics applied to image data, such as grain size, phase fraction and correlation coefficients.
This can provide some crucial information, but the metrics which are suitable in each case will depend on the morphology of the microstructure.
For example, interlamellar spacing is useful to describe lamellar microstructures, but is redundant when considering an equiaxed microstructure.
In any case, such measures embody only a tiny fraction of the information contained in a microstructural image.
A quantitative description of microstructure that is independent of morphology (a microstructural fingerprint) and embodies a full range of features would enable direct comparisons between microstructures with different morphologies, and provide new methods for constructing PSP relationships~\cite{withers2019, DECOST2015126}, or for quantifying the deviation of a given microstructure from some ideal standard in a quality control procedure.

In the past two decades, reasonable progress has been achieved in the quantification and low-dimensional representation of microstructures, over larger processing and material compositional windows~\cite{latypov2017, yabansu2017, popova2017}.
These advances have been possible with the use of concepts and toolsets from data science and informatics~\cite{rajan2015, mcdowell2016}.
The first potential benefit is realised with the use of automated feature engineering of the hierarchical material structures, e.g., establishing low-dimensional representations that provide maximum value in capturing high-fidelity PSP linkages.
A systematic and comprehensive quantification of the material structure is now possible, for example by performing statistical analysis of image features~\cite{white2023} or by combining the physics-inspired framework of $n$-point spatial correlations ($n$-point statistics) with machine learning approaches, such as principal component analysis (PCA)~\cite{fullwood2010}.

A machine learning approach that has not yet been fully explored, in the context of microstructural fingerprinting, is variational autoencoders (VAEs), which comprise a pair of convolutional neural networks (CNNs), referred to individually as the encoder and the decoder.
VAEs were first introduced by Kingma and Welling~\cite{kingma2014}.
Initial applications focussed on generating images of individual objects, particularly human faces, utilising datasets such as the CelebA dataset~\cite{liu2015faceattributes}.
More recently, attention has shifted towards machine learning applications in materials science, such as quantification of microstructure and the development of new PSP relationships.
The key difference is that microstructural image data are such that the entire image field contains potentially meaningful information, rather than an image of a foreground object of interest and some arbitrary background.
Attempts have been made to construct VAEs for texture embedding.
One example is TextureVAE, which consists of a VGG19 encoder and a decoder comprising 4 convolution blocks~\cite{sardeshmukh2021}.
The network was tested on various microstructures and latent dimensions were artificially varied to visualise their effect on the resulting reconstructions.
Further exploration of the encoded space has also been shown to carry the potential for material property prediction~\cite{cang2018, kim2021}.
Conversely, the ability to generate synthetic microstructures from specified properties was demonstrated~\cite{stein2020}.
Dimensionality reduction of the encoded space has also been shown to provide meaningful visualisation of the space, enabling property prediction from microstructural image data~\cite{pathak2020} and identification of new processing routes for target orientation distributions~\cite{sundar2020}.

Here, we employ a deep residual block VAE architecture, based on ResNet18~\cite{resnet}, to encode optical micrograph data from two classes of Ti-6Al-4V; a lamellar microstructure and a bimodal equiaxed microstructure.
The encoded space is explored via paths and Gaussian permutations, as a tool for generating artificial microstructures.
The latent space in further explored through dimensionality reduction via $t$-SNE, with analysis of morphological metric distributions across the space.
Support vector regression (SVR) is then applied to correlate fingerprints contained in the latent space with the same morphological metrics to quantify the trends visualised by the $t$-SNE\@.

Our titanium alloy dataset is representative of the microstructures of an important class of industrial materials in which control of microstructure is key to delivering the required in-service performance and in which quality assurance of material (and hence microstructure) is a critical part of manufacturing processes.
Furthermore, the nature of the dataset allows us to evaluate the performance of the VAE with respect to several aspects common to a broad range of metallurgical challenges:
\begin{enumerate}
    \item the descriptors should transparently handle a range of microstructures (here we have widely varying grain morphology);
    \item the space of descriptors (the latent space of the VAE) should be well-behaved, in the sense that the fingerprints should vary smoothly with changes to the microstructure and vice versa;
    \item the representation in the latent space should be interpretable in terms of key features of the microstructure (or properties of the material), which is to say that the fingerprints should show statistical correlation with features and properties.
\end{enumerate}
We explore these aspects of the VAE performance for our titanium dataset.


\section{Materials and Methods}
\label{sec:methods}

\subsection{Dataset}
\label{subsec:dataset}

\begin{figure}[!ht]
    \centering
    \begin{subfigure}{0.4\textwidth}
        \centering
        \includegraphics[width=\textwidth]{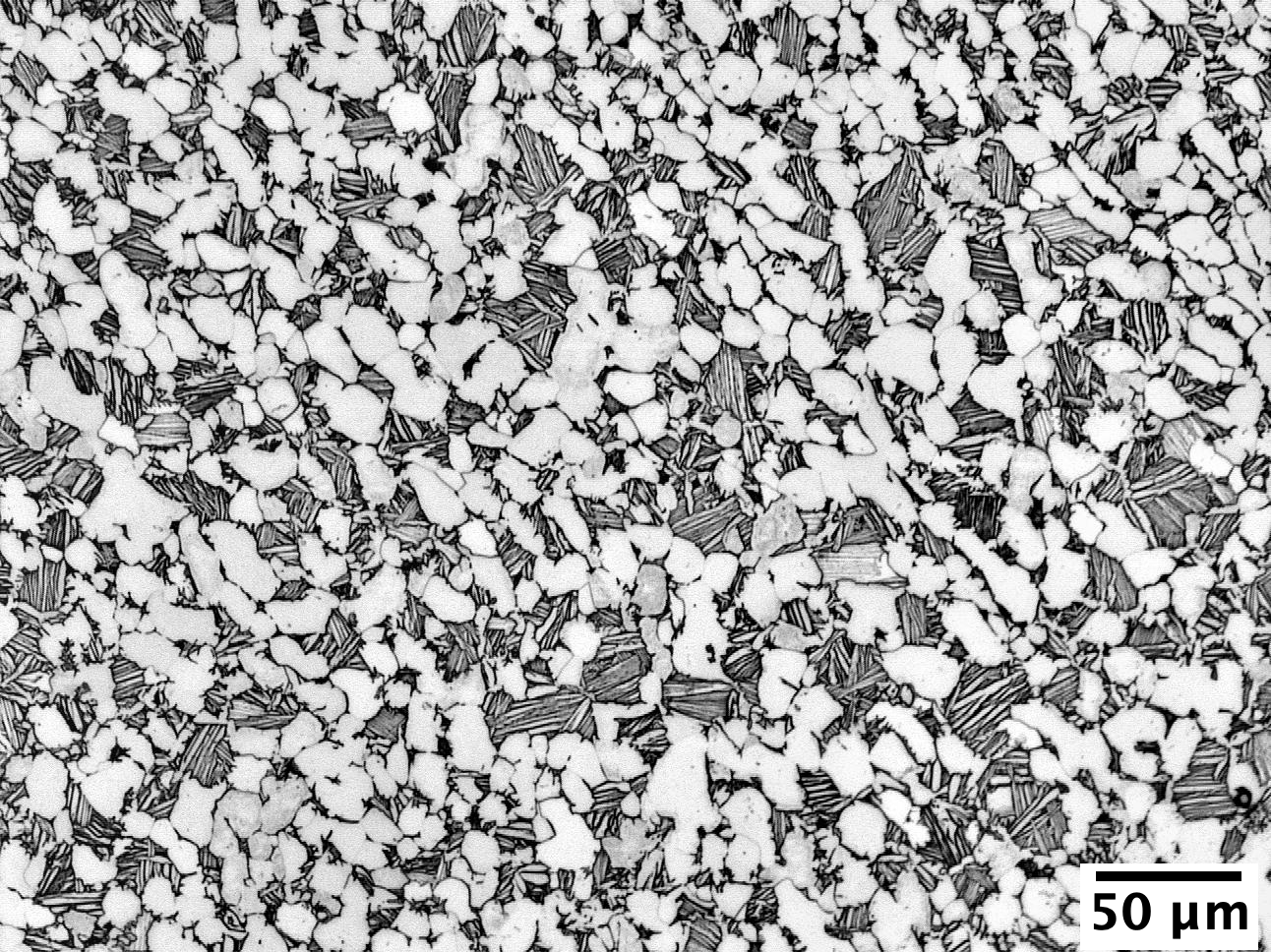}
        \caption{Bimodal}
    \end{subfigure}
    \hfill
    \begin{subfigure}{0.4\textwidth}
        \centering
        \includegraphics[width=\textwidth]{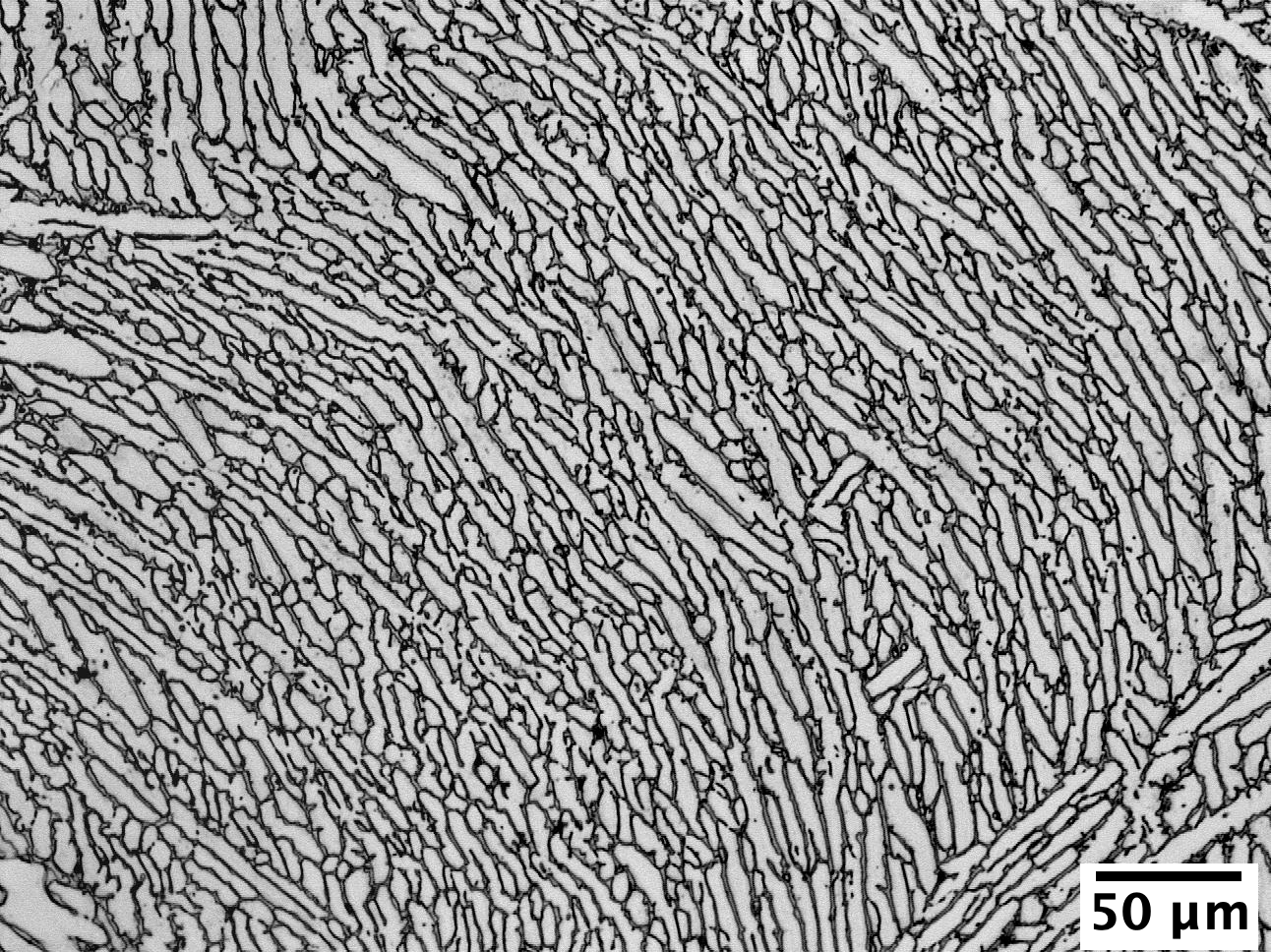}
        \caption{Lamellar}
    \end{subfigure}
    \caption{Representative optical micrographs of Ti-6Al-4V from LFTi64BL dataset.}
    \label{fig:lfti64bl}
\end{figure}

The LightForm Ti-6Al-4V alloy bimodal/lamellar dataset (LFTi64BL) is an open access dataset, curated within the LightForm group at the University of Manchester, and can be accessed via Zenodo~\cite{lfti64bl}.
The dataset comprises 40 optical 8 bit micrographs of resolution $1292 \times 968$ pixels, containing equal numbers of bimodal equiaxed and lamellar microstructures.
An example from each classification is shown in Figure~\ref{fig:lfti64bl}.

This dataset contains a relatively small number of images for machine learning purposes.
To expand the dataset, patches were extracted with random rotations and reflections applied.
The expanded LFTi64BL dataset contains 1,000 patches from each image, resulting in 40,000 patches of resolution $256 \times 256$ pixels.

\subsection{Greyscale Normalisation}
\label{subsec:greyscale}

\begin{figure}[!ht]
    \centering
    \begin{subfigure}{0.45\textwidth}
        \centering
        \includegraphics[width=\textwidth]{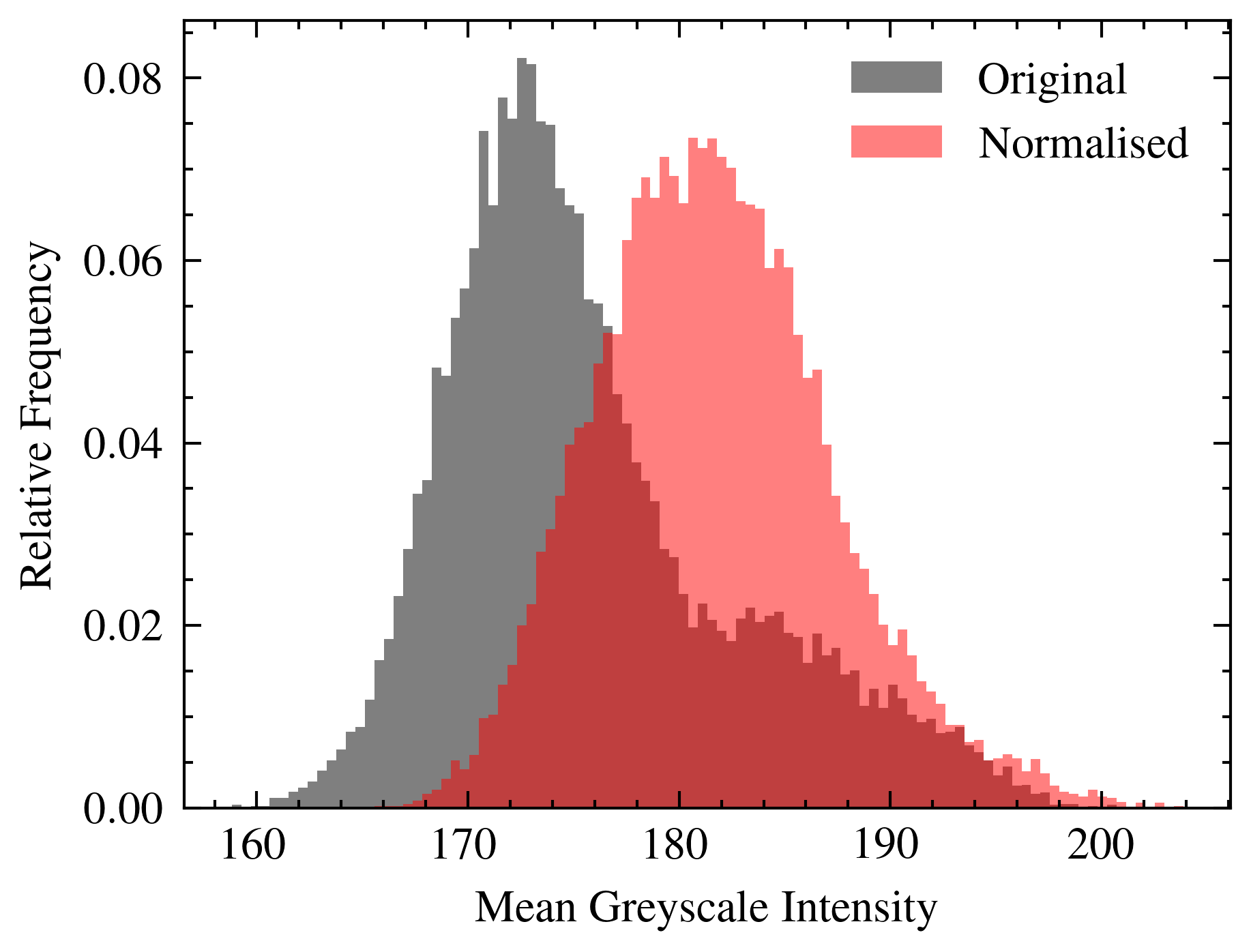}
        \caption{Bimodal}
    \end{subfigure}
    \hfill
    \begin{subfigure}{0.45\textwidth}
        \centering
        \includegraphics[width=\textwidth]{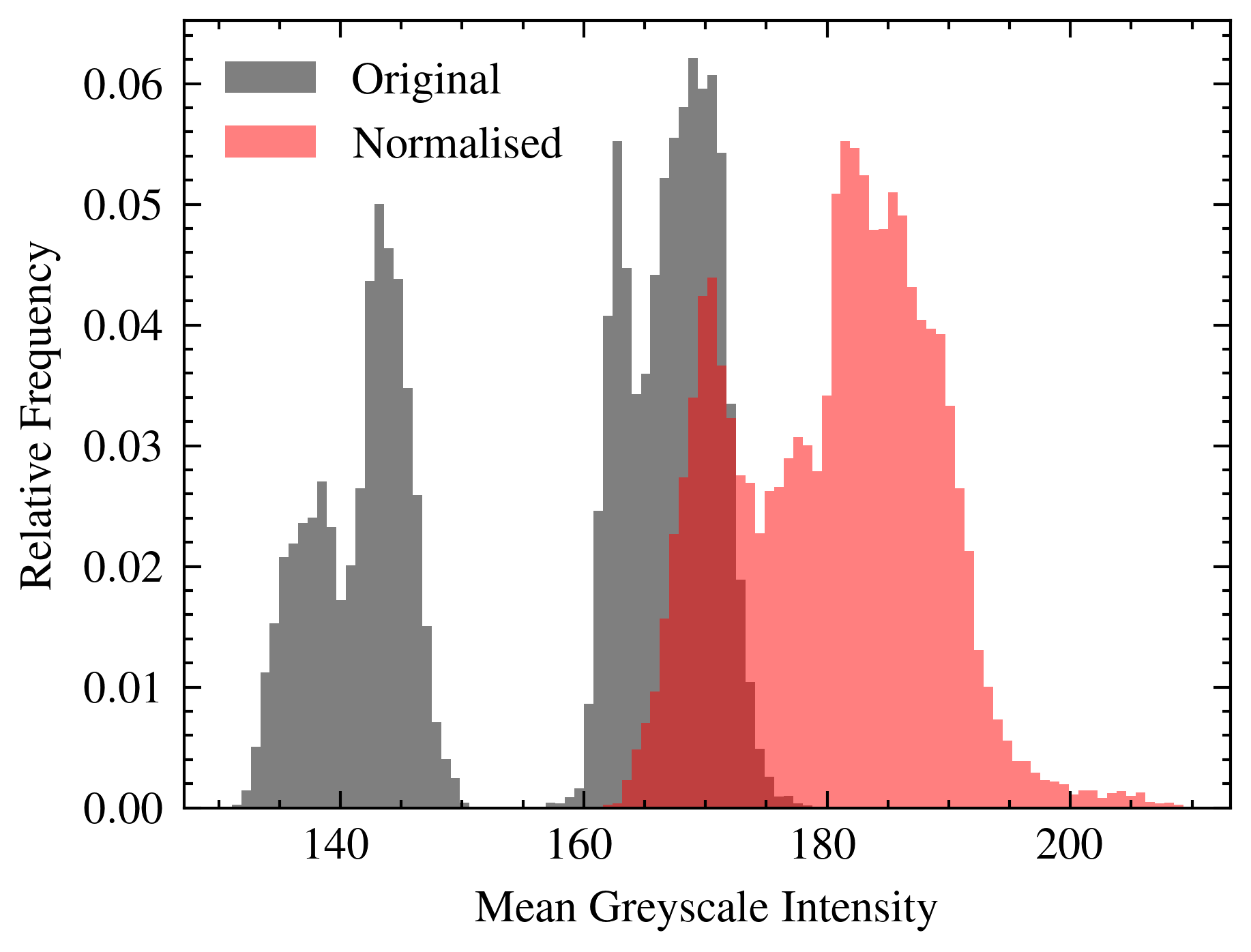}
        \caption{Lamellar}
    \end{subfigure}
    \caption{Distribution of mean greyscale intensities across the LFTi64BL dataset before and after normalisation, for bimdoal and lamellar microstructures separately.}
    \label{fig:greyscale_hist}
\end{figure}

Each image was white balanced by clipping greyscale intensities outside the 90$^{\text{th}}$ percentile and remapping to the range $[0, 1]$.
Figure~\ref{fig:greyscale_hist} shows distributions of mean greyscale intensities across the bimodal and lamellar datasets within LFTi64BL separately, before and after normalisation.

Prior to white-balancing, the variation in mean greyscale intensity is heavily influenced by fluctuations in lighting conditions during image capture.
However, after white-balancing, greyscale intensity is more normally distributed across the dataset.
This is now indicative of the distribution of phase fractions, with lower mean greyscale intensity corresponding to a higher volume fraction of the $\upbeta$ phase.

\subsection{Variational Autoencoders (VAEs)}
\label{subsec:vaes}

\begin{figure}[!ht]
    \centering
    \includegraphics[width=.8\textwidth]{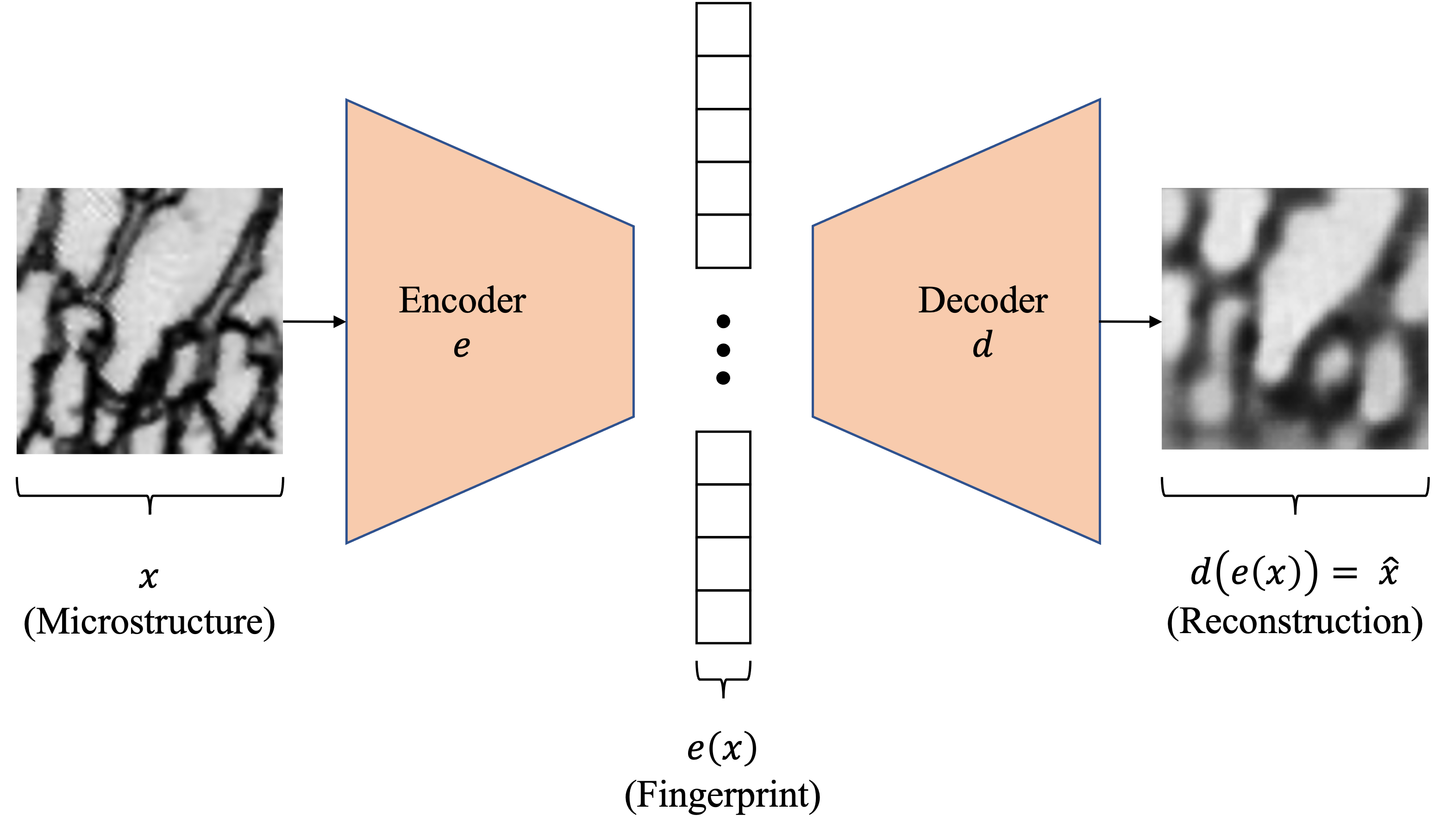}
    \caption{Schematic of general VAE architecture that takes an image $x$ as input, encodes the image into a fingerprint $e(x)$ and is trained to reconstruct the input image with the mapping $d(e(x))$.}
    \label{fig:vae_schematic}
\end{figure}

Variational autoencoders (VAEs) are a tool for encoding image data into a compressed format (or fingerprint) that preserves morphological features.
They comprise a pair of convolutional neural networks (CNNs), referred to individually as the encoder and decoder.
Fingerprints output by the trained encoder can be fed into downstream tasks such as image classification and property prediction.
Figure~\ref{fig:vae_schematic} provides a visual interpretation of the general VAE architecture.

Suppose we have an image, $x \in \mathbb{R}^{m_1 \times m_2}$.
The encoder takes the image $x$ as an input and computes a compressed representation $z \in \mathbb{R}^D$, such that $D \ll m_1 m_2$, where $z = e(x)$ and $D$ is a tunable parameter that denoted the dimensionality of the encoded space.
The decoder then takes the encoded vector $z$ as an input and aims to reconstruct the input image, $x$.
We denote the reconstruction as $\hat{x} = d(e(x))$, where $\hat{x} \in \mathbb{R}^{m_1 \times m_2}$.
The encoded space learned is continuous and normally distributed.

To train a VAE, we must have a set of images, $X = \{x_1, x_2, \dots, x_N\}$, where $N$ is the total number of images in the dataset.
The dataset is split into two subsets, $X_{\text{train}} \in \mathbb{R}^{N_{\text{train}} \times m_1 \times m_2}$ and $X_{\text{eval}} \in \mathbb{R}^{N_{\text{eval}} \times m_1 \times m_2}$, where $X_{\text{train}}$ is used to train the VAE, $X_{\text{eval}}$ is used for evaluating the VAE on unseen data, $N_{\text{train}} + N_{\text{eval}} = N$, $X_{\text{train}} \cup X_{\text{eval}} \equiv X$ and $X_{\text{train}} \cap X_{\text{eval}} \equiv \varnothing$.
The training set is split into batches and fed into the VAE, one batch at a time.
Ultimately, $d(e(x_i))$ for $x_i \in X_{\text{train}}$ is computed and some loss function is used as a metric for assessing the encoded representation and reconstruction quality.
The loss function is then used as an input to an optimiser.
The Adam optimiser~\cite{ADAM} is currently the state of the art and is utilised throughout all models discussed herein.
The optimiser updates weights and biases in both the encoder and decoder from a single loss function after each batch operation.
The model can then be evaluated on $X_{\text{eval}}$ to measure the potential for transfer learning, but metrics calculated on these images in the evaluation set are not utilised for updating any weights or biases.
Encoded representations, generated from the trained VAE on microstructural image data, can be considered as a signature, or microstructural fingerprint (as described in~\cite{withers2019}), and will be referred to as such throughout this paper.

\subsection{Loss Functions}
\label{subsec:loss}

During VAE training, a loss function is periodically computed on the model, to determine performance and provide the inputs for the optimiser to update the weights and biases in the networks, with the aim of minimising the loss function.
This is typically a combination between a measure in the spacial domain of the reconstruction accuracy and imposing a restriction on the encoded space to be normally distributed~\cite{kingma2014}.
To measure how normally distributed the encoded space is, the Kullback-Leibler (KL) divergence, $D_{\text{KL}}$, can be determined between each encoded representation and the unit normal distribution $\mathcal{N}(0, 1)$~\cite{kingma2019}.

The reconstruction accuracy can be measured in several ways.
A popular method is the mean squared error (MSE)~\cite{yu2020}, which is a distance metric in the spacial domain, given by
\[
    \text{MSE} = \frac{1}{m} \sum_{j=1}^{m} (x_j - \hat{x}_j)^2,
\]
where $m = m_1 m_2$ is the total number of pixels in the image and the $x_j, \hat{x}_j$ denote pixels in the input and reconstruction, respectively.

Another metric that operates in the spacial domain is the binary cross-entropy (BCE), denoted here as $\mathcal{L}_b$, which is a measure of negative log likelihood and is given by
\[
    \mathcal{L}_b = - \sum_{i=1}^{m} x_i \log \hat{x}_i - \sum_{i=1}^{m} (1 - x_i) \log (1 - \hat{x}_i).
\]

The issue with minimising these loss functions is that they generally result in blurry reconstructions.
With the aim of minimising blur, a loss function on the frequency domain was proposed in~\cite{bjork2022}.
This requires calculating the 2D Fourier transform of both the input and reconstruction.
The spectral loss, $\mathcal{L}_f(x, \hat{x})$, can then be given by the MSE between the 2D Fourier transforms, i.e.,
\[
    \mathcal{L}_f(x, \hat{x}) = \frac{1}{m} \sum_{j=1}^{m} \left((\text{Im}\{\mathcal{F}(x)_j\} - \text{Im}\{\mathcal{F}(\hat{x})_j\})^2 + (\text{Re}\{\mathcal{F}(x)_j\} - \text{Re}\{\mathcal{F}(\hat{x})_j\})^2 \right),
\]
where $\mathcal{F}$ denotes the 2D FFT, $\text{Im}\{\mathcal{F}\}$ denotes the imaginary part of $\mathcal{F}$ and $\text{Re}\{\mathcal{F}\}$ the real part.
The total loss function to be minimised by the optimiser is then given by
\begin{equation}\label{eq:spec_loss}
\mathcal{L} = \alpha \mathcal{L}_b(x, \hat{x}) + (1 - \alpha) \mathcal{L}_f(x, \hat{x}),
\end{equation}
where $\alpha \in [0, 1]$ is a tunable hyperparameter.

\subsection{ResNet18-VAE}
\label{subsec:resnetvae}

ResNet~\cite{resnet} is a deep CNN composed of residual blocks and was initially proposed for classification of the ImageNet dataset~\cite{deng2009imagenet}, which is a dataset containing over 1 million natural images with 1000 classifications.
As such, the final layer of a standard ResNet is a 1000-dimensional fully connected layer, where the output from each node corresponds to a probability for each ImageNet classification.
The depth of the network can be controlled by varying the number of layers within each block and the total number of blocks.
Here, we consider ResNet18, which contains 8 residual blocks, each comprised of 2 convolution layers with subsequent ReLU activation and batch normalisation.
Each block has the potential to be effectively skipped by summing the input received at each block with the output from that block, after convolution.

\begin{figure}[!ht]
    \centering
    \begin{subfigure}{0.45\textwidth}
        \centering
        \includegraphics[width=\textwidth]{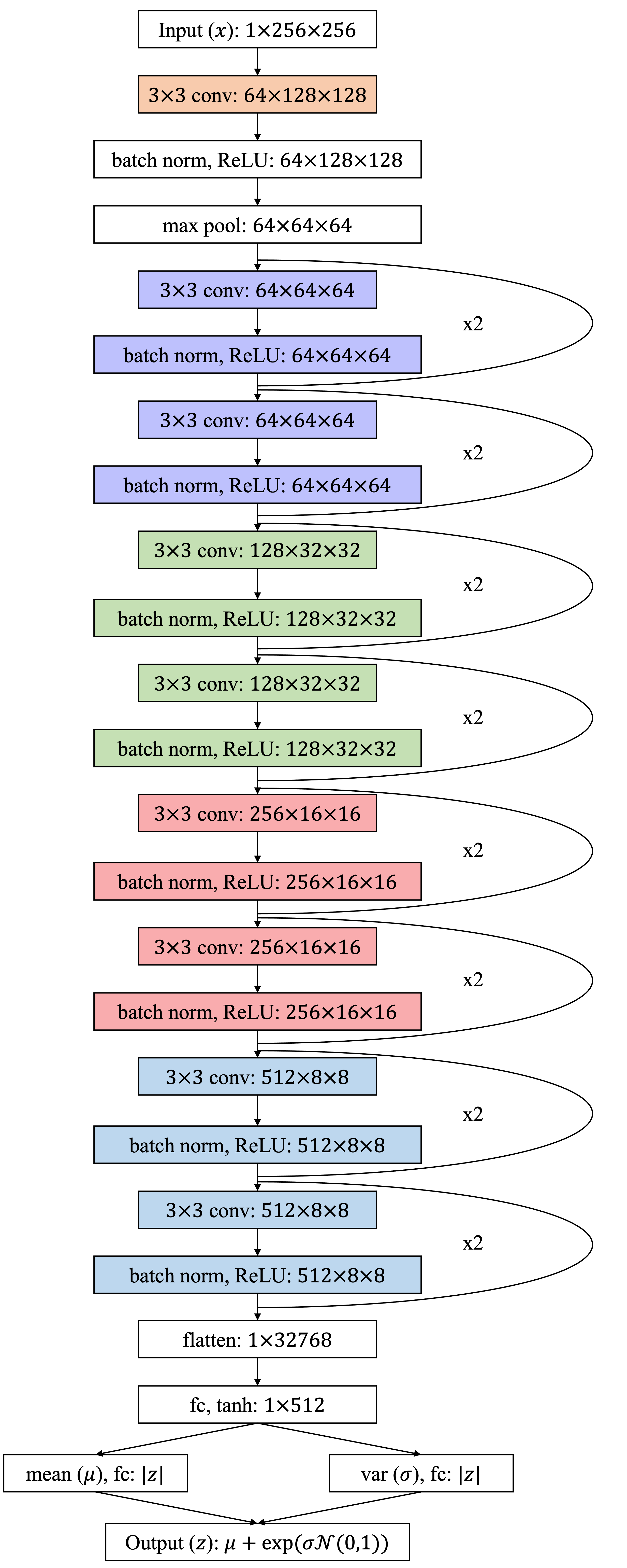}
        \caption{Encoder}
    \end{subfigure}
    \hfill
    \begin{subfigure}{0.45\textwidth}
        \centering
        \includegraphics[width=\textwidth]{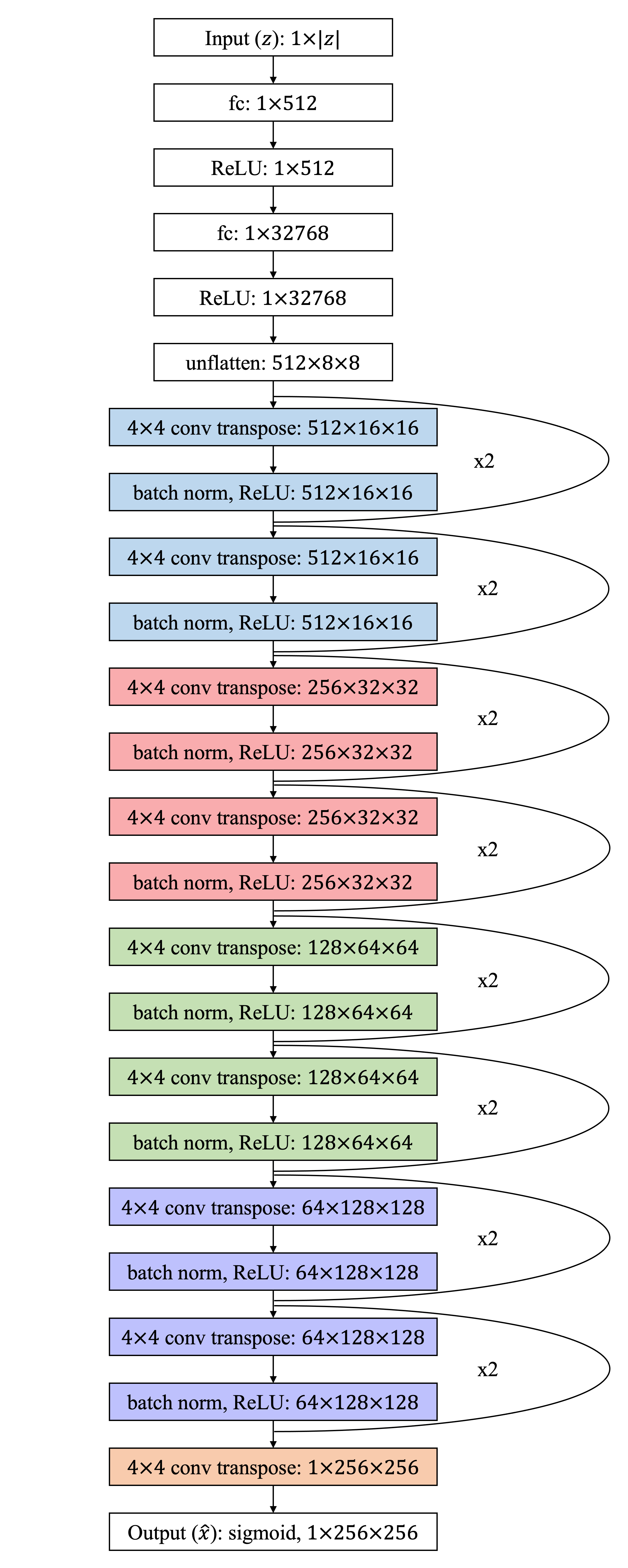}
        \caption{Decoder}
    \end{subfigure}
    \caption{ResNet18 VAE architecture for the $256 \times 256$ inputs used in the present study.}
    \label{fig:resnet_vae}
\end{figure}

To convert ResNet18 into an encoder, it was modified by replacing the average pool and 1000-dimensional fully connected layers at the end of the network with a flattening of the final convolution output.
This was followed by a 512-dimensional fully connected layer with tanh activation, which is then simultaneously fed into two separate fully connected layers.
Each fully connected layer consists of $|z|$ neurons, where $|z|$ is the specified dimension of the encoded space.
This is somewhat arbitrary, but 256 was used to generate the results presented in this paper.
One of these fully connected layers is utilised as a set of means, $\mu$, whilst the other is treated as a set of standard deviations, $\sigma$.
These are then combined into the output $z$ as
\[
    z = \mu + \exp(\sigma \mathcal{N}(0, 1)),
\]
where $\mathcal{N}(0, 1)$ denotes the standard normal distribution, with mean 0 and standard deviation 1.
This branching into $\mu$ and $\sigma$, with subsequent combination of the two, is what sets variational autoencoders apart from standard autoencoders.
The decoder then essentially mirrors the encoder with transpose convolution layers replacing the standard convolution layers.
The architecture for this ResNet18 VAE is provided in Figure~\ref{fig:resnet_vae}.

\subsection{Morphological Analysis}
\label{subsec:morphology}

Morphological measurements were automated for the LFTi64BL dataset as metrics for correlation with fingerprints produced by the VAE\@.
These metrics can then also be plotted as colour maps over across a dimensionality reduction of the fingerprint space to visualise how such features are distributed.
Each metric requires the images to be binarised prior to measurement.

\subsubsection{Image Binarisation}
\label{subsubsec:binarise}

\begin{figure}[!ht]
    \centering
    \begin{subfigure}{0.225\textwidth}
        \centering
        \includegraphics[width=\textwidth]{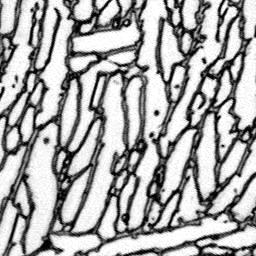}
        \caption{Original lamellar}
    \end{subfigure}
    \hfill
    \begin{subfigure}{0.225\textwidth}
        \centering
        \includegraphics[width=\textwidth]{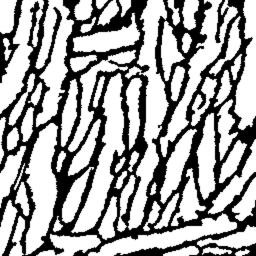}
        \caption{Binary lamellar}
    \end{subfigure}
    \hfill
    \begin{subfigure}{0.225\textwidth}
        \centering
        \includegraphics[width=\textwidth]{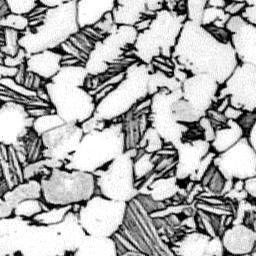}
        \caption{Original bimodal}
    \end{subfigure}
    \hfill
    \begin{subfigure}{0.225\textwidth}
        \centering
        \includegraphics[width=\textwidth]{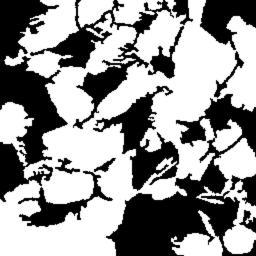}
        \caption{Binary bimodal}
    \end{subfigure}
    \caption{Representative optical micrograph images patches of Ti-6Al-4V before and after binarisation.}
    \label{fig:binary}
\end{figure}

Binarised images are required to compute certain metrics on the images, such as phase fraction and grain size.
A high-pass Gaussian filter was applied to each image, in the Fourier domain, to normalise illumination across the images.
The images were then binarised with Otsu's thresholding method~\cite{Bangare2015}, before applying area closing to remove noise from all images and $\upalpha_{\text{s}}$ laths from the bimodal images.
Pixels corresponding to $\upalpha_{\text{p}}$ grains are assigned a label, whilst all other pixels are labelled as 0.
Figure~\ref{fig:binary} provides representative examples of the resulting binary images.
The binarised lamellar microstructures are highly accurate, but there is some retention of $\upalpha_{\text{s}}$ laths in the bimodal microstructures due to connectivity with the $\upalpha_{\text{p}}$ grains.

\subsubsection{Phase Fraction}
\label{subsubsec:phase_frac}

Due to the way in which the images were binarised (described in Section~\ref{subsubsec:binarise}), the phase fraction is simply determined as the sum of all the pixel values in the binary image divided by the total number of pixels in the image.

\subsubsection{Lamellae Direction}
\label{subsubsec:direction}

\begin{figure}[!ht]
    \centering
    \begin{subfigure}{0.225\textwidth}
        \centering
        \includegraphics[width=\textwidth]{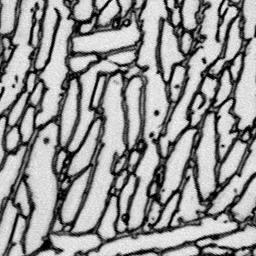}
        \caption{Original micrograph}
    \end{subfigure}
    \hfill
    \begin{subfigure}{0.225\textwidth}
        \centering
        \includegraphics[width=\textwidth]{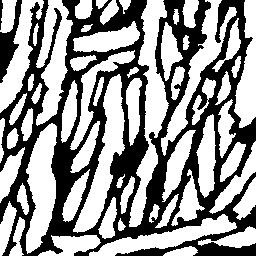}
        \caption{Binary with area closing}
    \end{subfigure}
    \hfill
    \begin{subfigure}{0.225\textwidth}
        \centering
        \includegraphics[width=\textwidth]{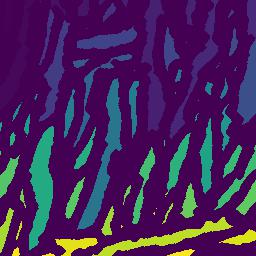}
        \caption{Eroded and labelled}
    \end{subfigure}
    \hfill
    \begin{subfigure}{0.225\textwidth}
        \centering
        \includegraphics[width=\textwidth]{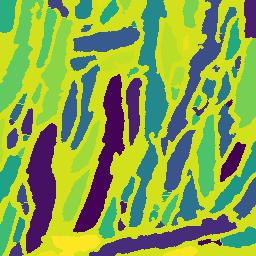}
        \caption{Watershed segmentation}
    \end{subfigure}
    \caption{Erosion and watershed methods applied to a representative micrograph for grain isolation.}
    \label{fig:watershed}
\end{figure}

\begin{figure}[!ht]
    \centering
    \includegraphics[width=0.45\textwidth]{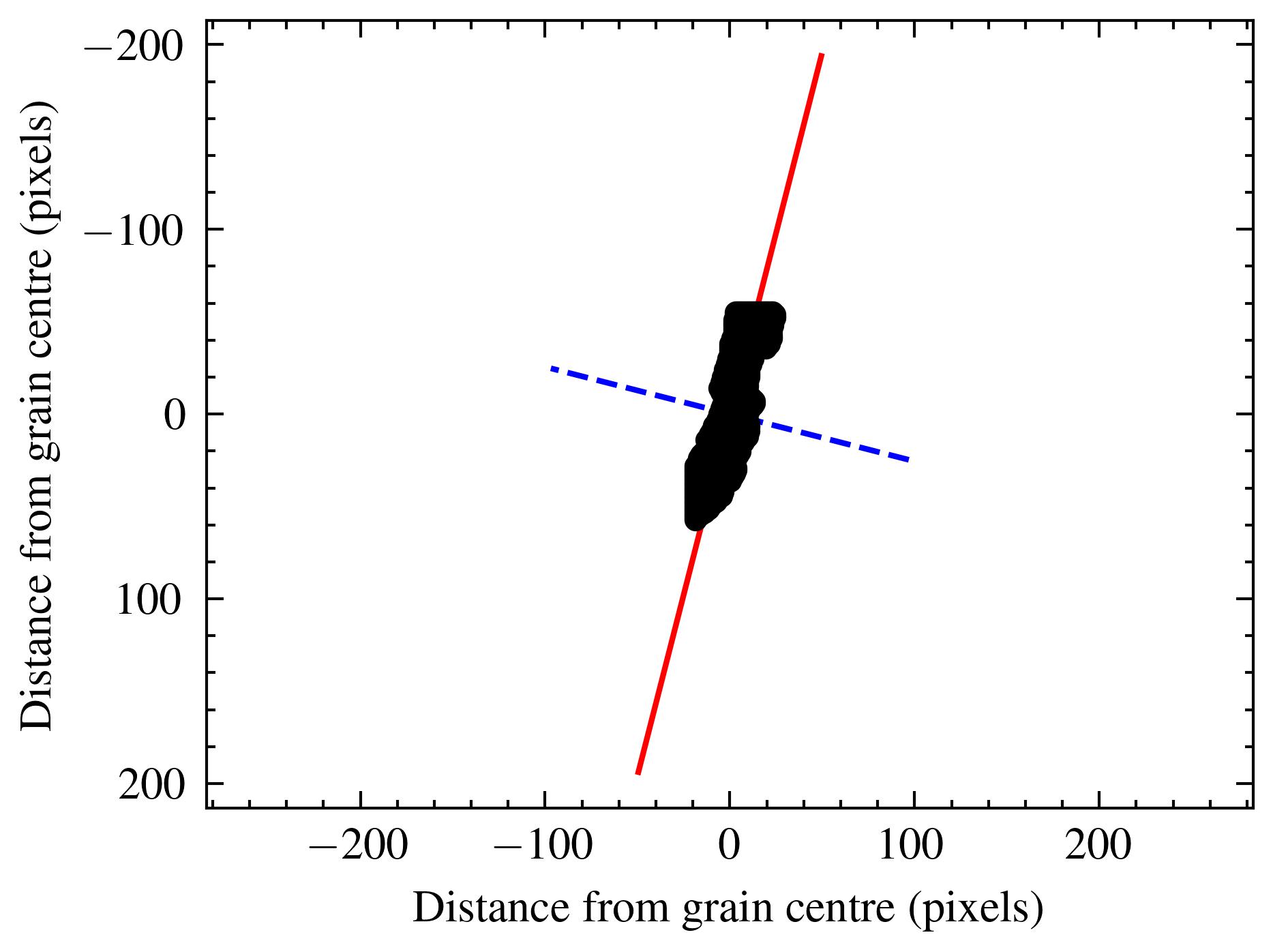}
    \caption{Eigenvectors plotted on an individual lamella for direction measurement, with the primary eigenvector shown in red and the secondary eigenvector plotted in blue.}
    \label{fig:eigen}
\end{figure}

For the lamellar images in the LFTi64BL dataset, another descriptor we can consider is the dominant direction, or orientation, of the lamellae, relative to the bottom edge of the image.
This can be quantified as the mean angle subtended between the elongation direction of each lamella and the bottom edge of the image.
An erosion is applied on the binary image to isolate overlapping grains before each grain is labelled with a unique integer.
Watershed segmentation is then applied with markers taken from the labelled image and the initial binary image as a mask.
Each grain in the watershed image is isolated and eigenvectors are computed for each grain.
The primary eigenvector describes the direction of each lamella.
The angle between the primary eigenvector and the bottom edge of the image is then calculated for each grain and averaged to provide the metric for lamellae direction of a given image.
Figure~\ref{fig:watershed} shows the erosion and watershed method applied to a representative micrograph and Figure~\ref{fig:eigen} shows the corresponding eigenvectors for an individual grain.

\subsubsection{Bimodal Grain Size}
\label{subsubsec:grain_size}

As a metric for quantifying the bimodal microstructures in LFTi64BL, average grain size measurements were automated following ASTM E1382-97~\cite{astm_e1382}.
Random line scans are applied on the binarised bimodal micrographs and peaks in the derivative of the profile along the line scans are used to detect grain edges.
The distance in pixels between peaks in the derivative of the profile line is then converted into a distance in $\mu$m.

\subsection{Support Vector Regression (SVR)}
\label{subsec:svr}

Once fingerprints have been extracted from the ResNet18 VAE, regression algorithms can be trained to predict morphological metrics.
Here, we use support vector regression (SVR), which is an extension to support vector machines for continuous variables~\cite{Awad2015}.
Gaussian process regression is a popular alternative to SVR, but relies on relatively low-dimensional inputs compared to the dimensionality of the ResNet18 VAE fingerprints.
Fingerprints are randomly split into a training set containing 90\% of the fingerprints and a test set containing 10\% for input into the SVR\@.
This is repeated 10 times with random splits to perform 10-fold cross-validation.

\subsection{$t$-Stochastic Neighbour Embedding ($t$-SNE)}
\label{subsec:tsne}

The encoded space learned is high-dimensional (256 dimensions for ResNet18-VAE architecture, outlined in Section~\ref{subsec:resnetvae}).
This makes it difficult to visualise the encoded space in its entirety.
To aid in visualising this high-dimensional space, we can perform dimensionality reduction down to 2 or 3 dimensions, which allows us to plot the encoded space and to assess how the microstructures in our training set are distributed.
Here, we consider $t$-stochastic neighbour embedding ($t$-SNE)~\cite{silva_tsne_2023}.

The fingerprints are represented as a similarity matrix, $S$, where entries, $S_{i, j}$, denote the probability that $z_j$ is a nearest neighbour of $z_i$.
The aim is then to minimise the distance between $S_{i, j}$ and $S_{j, i}$~\cite{hinton2002}.
Standard SNE utilises a Gaussian distribution to determine similarity between fingerprints, whereas $t$-SNE relies on the Student's $t$ distribution, hence its name.
Due to the longer tail of the $t$ distribution, relative to a Gaussian distribution, the use of the $t$ distribution results in more separated embeddings and helps alleviate the crowding issue often encountered with SNE~\cite{vandermaaten2008}.
Principal component analysis (PCA) is used to initialise the $t$-SNE\@.

\subsection{Traversing the Encoded Space}
\label{subsec:traversing}

One way in which the encoded space can be traversed is through construction of a specific path.
Here, we consider a linear path.
Two images, $x_1$ and $x_2$, are randomly selected and $e(x_1), e(x_2)$ are computed.
A linear path is then constructed from $e(x_1)$ to $e(x_2)$ in the encoded space, according to the following equation.
\[
    e_n(x_1, x_2) = e(x_1) + \frac{n + 1}{N}(e(x_2)), \hspace{1em} n = 1, \dots, N,
\]
where $e_n(x_1, x_2)$ denotes each fingerprint along the path and N is the number of steps along the path.
Fingerprints along the path are then supplied to the decoder to output reconstructions.

Another method for generating potentially valid encoded representations of microstructure is to form a random cloud, centered at a known valid encoded representation, $e(x)$, for some image $x$.
This can be achieved by sampling random Gaussian noise from $\mathcal{N}(0, 1)$ and summing with $e(x)$, i.e.,
\[
    e_n(x) = e(x) + \gamma \mathcal{N}(\mu, \sigma), \hspace{1em} n = 1, \dots, N,
\]
where $e_n(x)$ is a random neighbour of $e(x)$, $\gamma$ is a tunable scale factor that controls the noise level, $\mu, \sigma \in \mathbb{R}^D$ denote the element-wise mean and standard deviation for each dimension in the encoded space and $N$ is the number of neighbouring fingerprints to be output.


\section{Results}
\label{sec:results}

\subsection{Reconstruction Accuracy}
\label{subsec:recon_acc}

\begin{figure}[!ht]
    \centering
    \includegraphics[width=\textwidth]{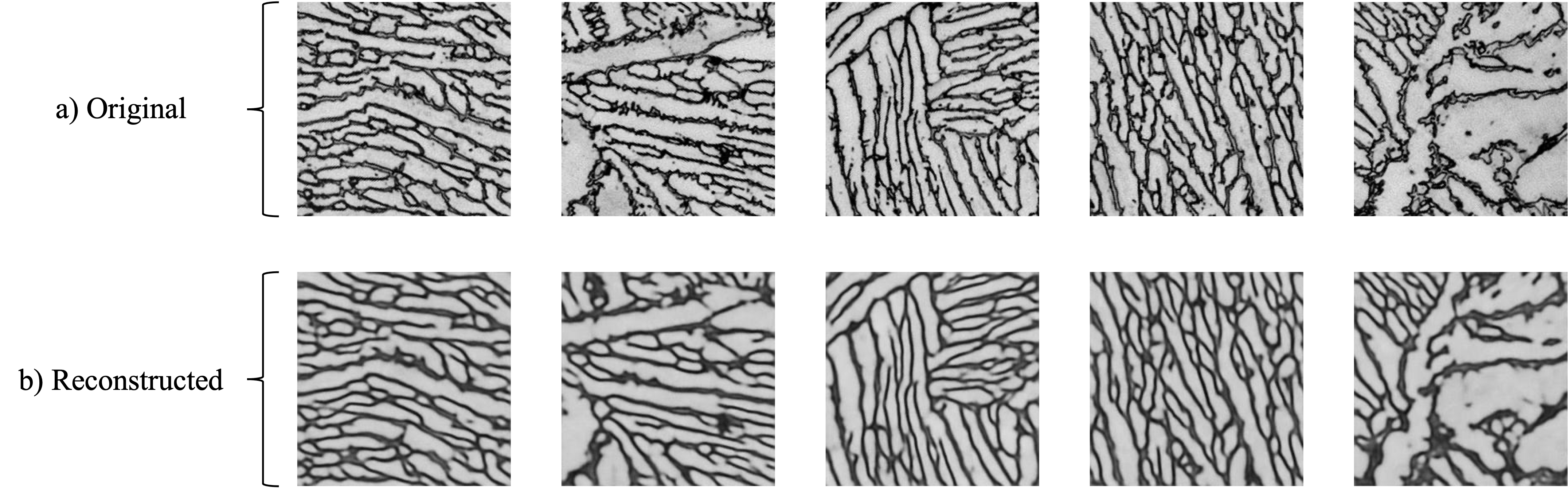}
    \caption{Representative examples of (a) $256 \times 256$ patches from the lamellar microstructures and (b) their corresponding reconstructions from the ResNet18 VAE architecture, after 1000 epochs.}
    \label{fig:resnet_recons_lamellar}
\end{figure}

\begin{figure}[!ht]
    \centering
    \begin{subfigure}{0.45\textwidth}
        \centering
        \includegraphics[width=\textwidth]{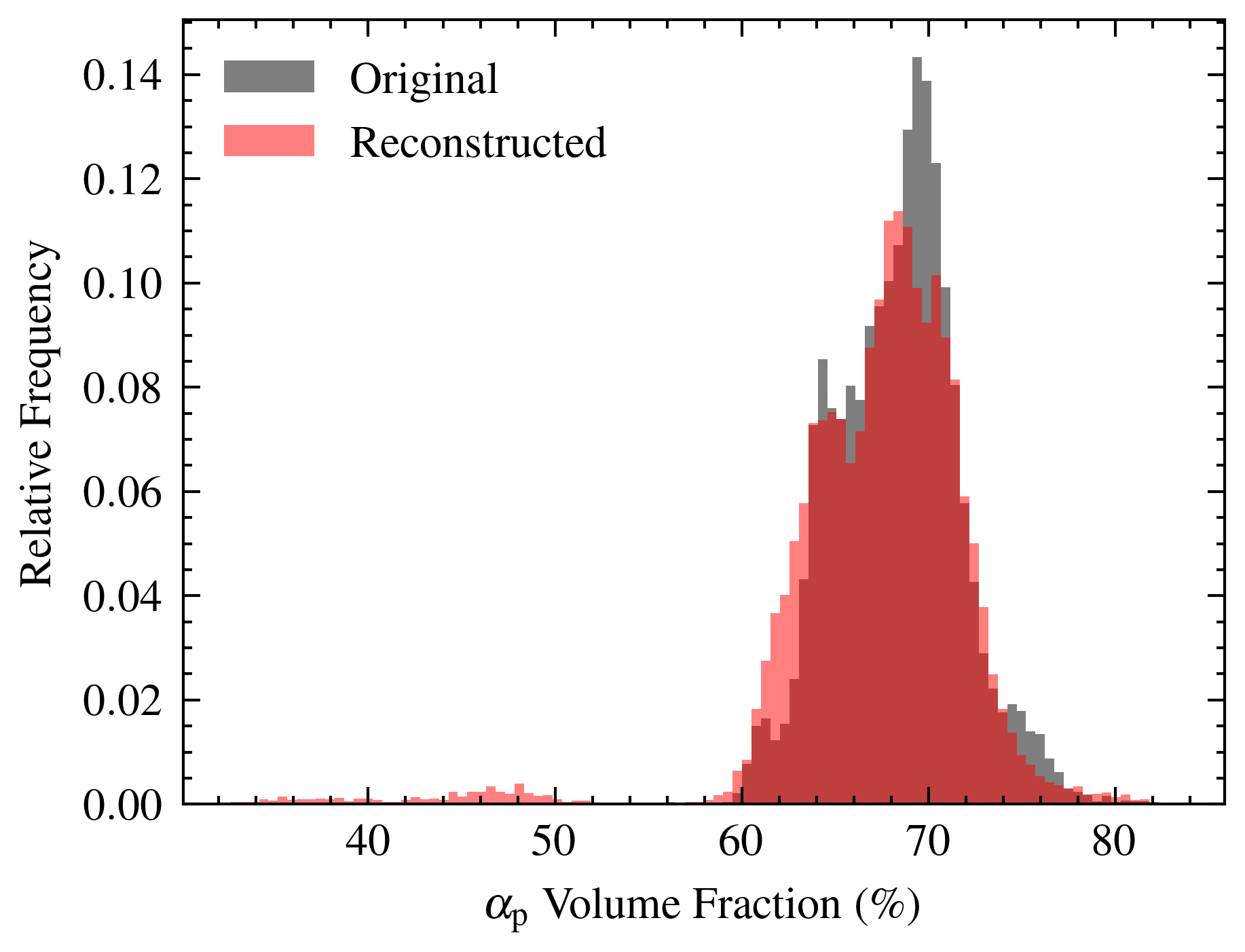}
        \label{subfig:vol_frac_lamellar}
    \end{subfigure}%
    \hfill
    \begin{subfigure}{0.45\textwidth}
        \centering
        \includegraphics[width=\textwidth]{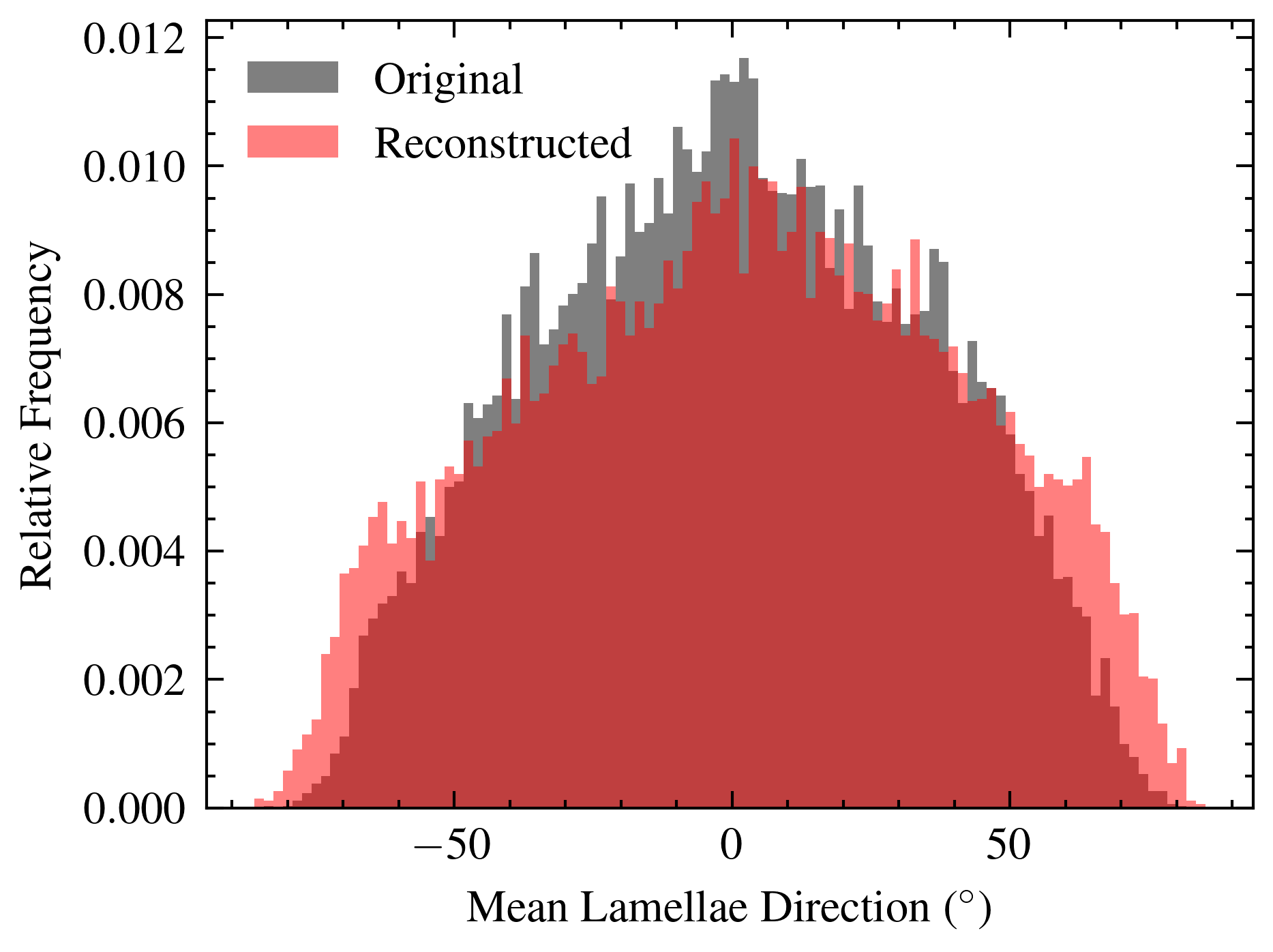}
        \label{subfig:direction_lamellar}
    \end{subfigure}
    \centering
    \begin{subfigure}{0.45\textwidth}
        \centering
        \includegraphics[width=\textwidth]{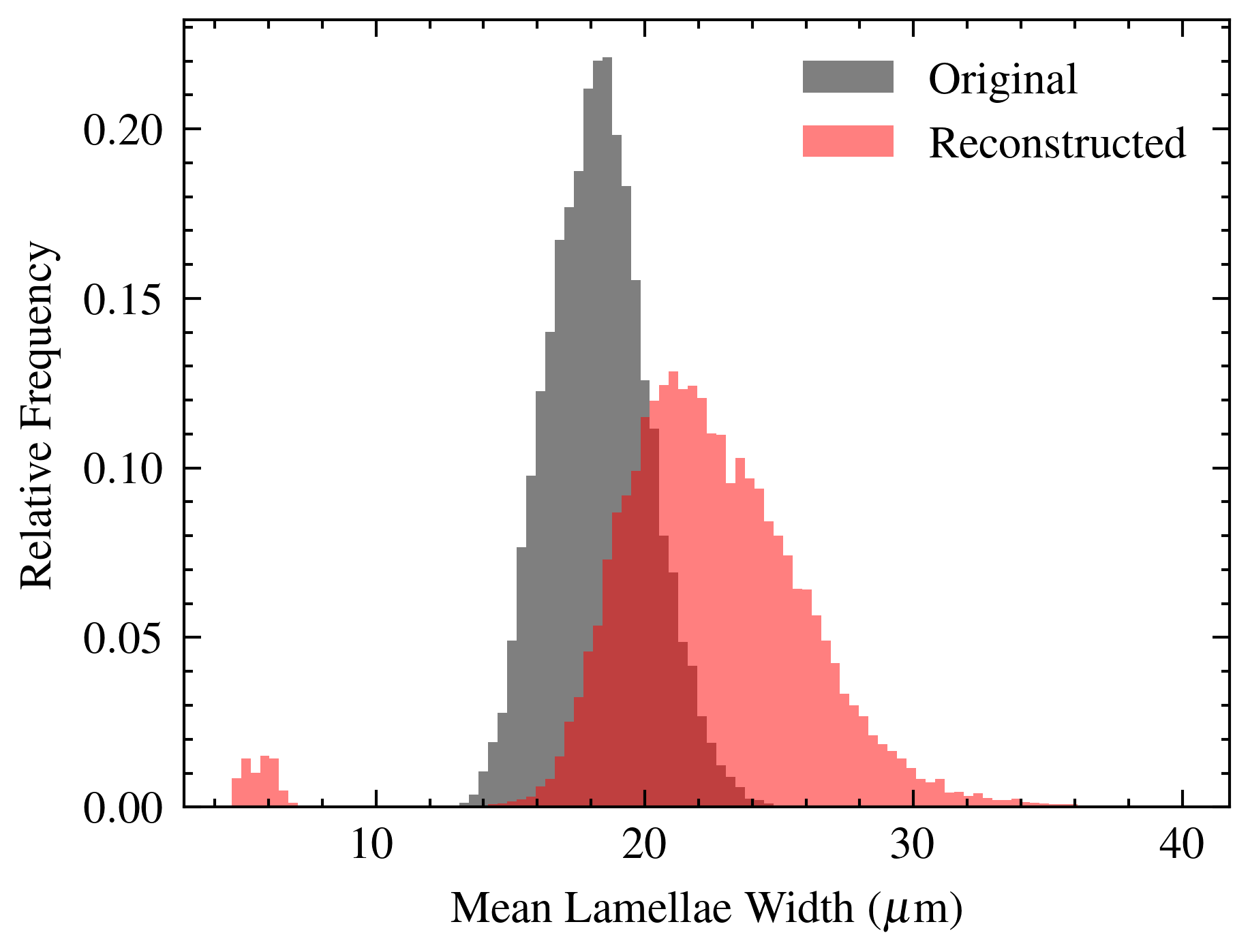}
        \label{subfig:width_lamellar}
    \end{subfigure}%
    \hfill
    \begin{subfigure}{0.45\textwidth}
        \centering
        \includegraphics[width=\textwidth]{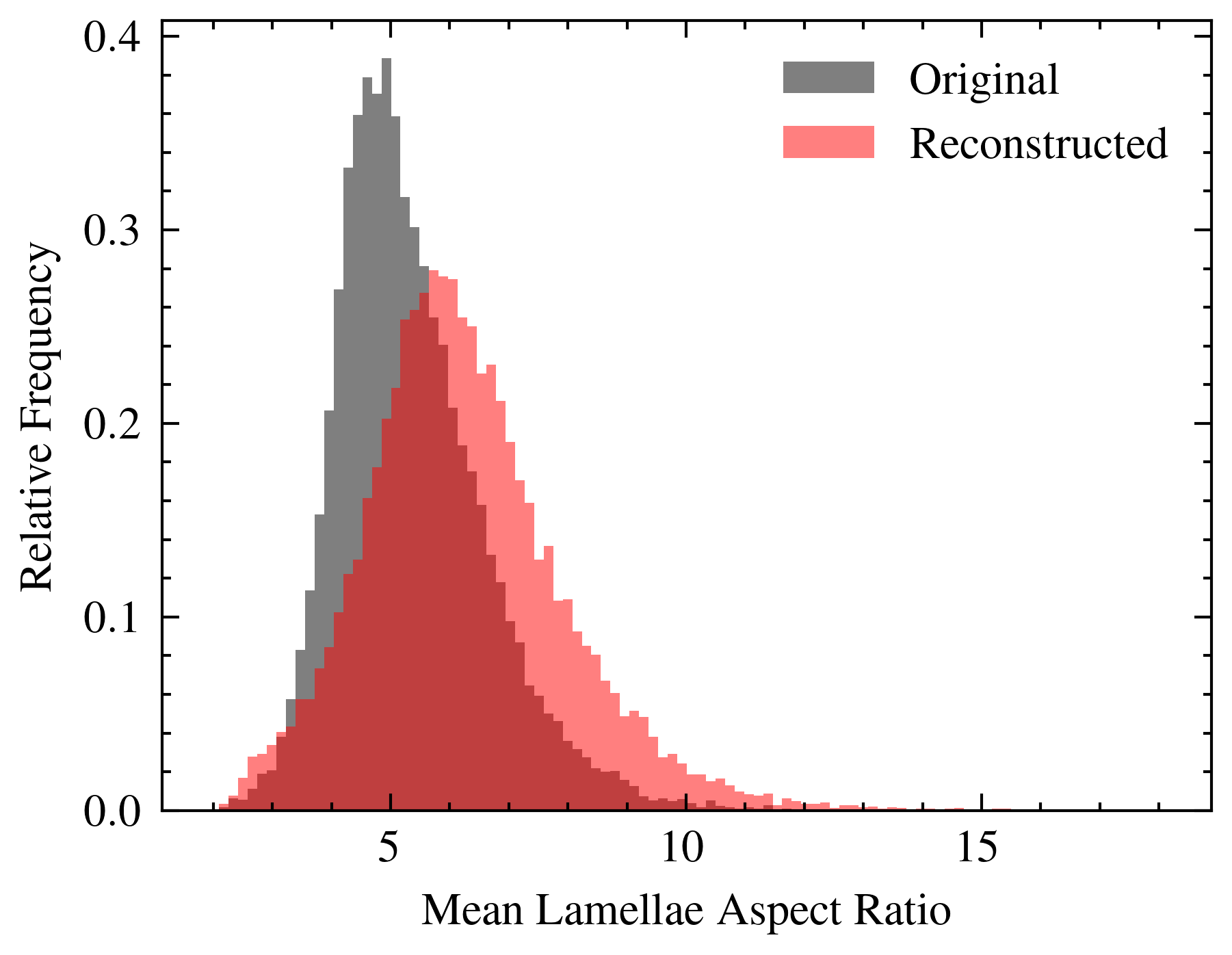}
        \label{subfig:aspect_lamellar}
    \end{subfigure}
    \caption{Morphological metric distributions across the original image patches and their corresponding reconstructions for the lamellar dataset.}
    \label{fig:metrics_lamellar}
\end{figure}

The ResNet18 VAE was trained under two separate regimes.
The first included exclusively either lamellar or bimodal microstructures and the second included the full LFTi64BL dataset, containing both lamellar and bimodal microstructures.
Figure~\ref{fig:resnet_recons_lamellar} shows some example reconstructions from the lamellar dataset after 1000 epochs.
When training is restricted to the lamellar dataset, grain boundaries are accurately identified and reconstructed, but there is some smoothing evident in the reconstructions and the interlamellar $\upbeta$ appears coarsened.
This becomes clearer upon inspection of the morphological metric distributions shown in Figure~\ref{fig:metrics_lamellar}.
The $\upalpha_\text{p}$ volume fraction is slightly reduced in the reconstructions, with some anomalies between 30\% and 50\%.
The mean lamellae direction, relative to the bottom edge of each image patch, is consistent.
This confirms that the $\upalpha_{\text{p}}$ grains are oriented correctly in the reconstructions, but there is an increase in the mean lamellae width, which is likely due to the smoothing effect removing smaller grains.
There is also an increase in the mean lamellae aspect ratio, owed to the coarsening of the interlamellar $\upbeta$ in the reconstructions which results in $\upalpha_{\text{p}}$ grains appearing more elongated.

\begin{figure}[!ht]
    \centering
    \includegraphics[width=\textwidth]{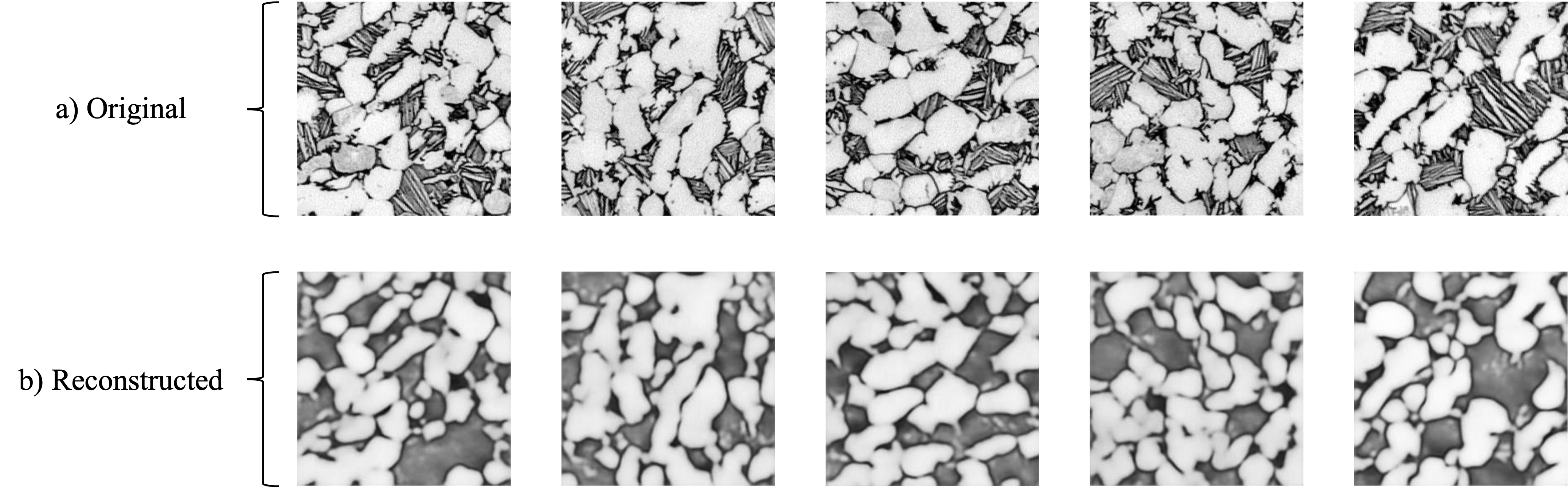}
    \caption{Representative examples of (a) $256 \times 256$ patches from the bimodal microstructures and (b) their corresponding reconstructions from the ResNet18 VAE architecture, after 1000 epochs.}
    \label{fig:resnet_recons_bimodal}
\end{figure}

\begin{figure}[!ht]
    \centering
    \begin{subfigure}{0.45\textwidth}
        \centering
        \includegraphics[width=\textwidth]{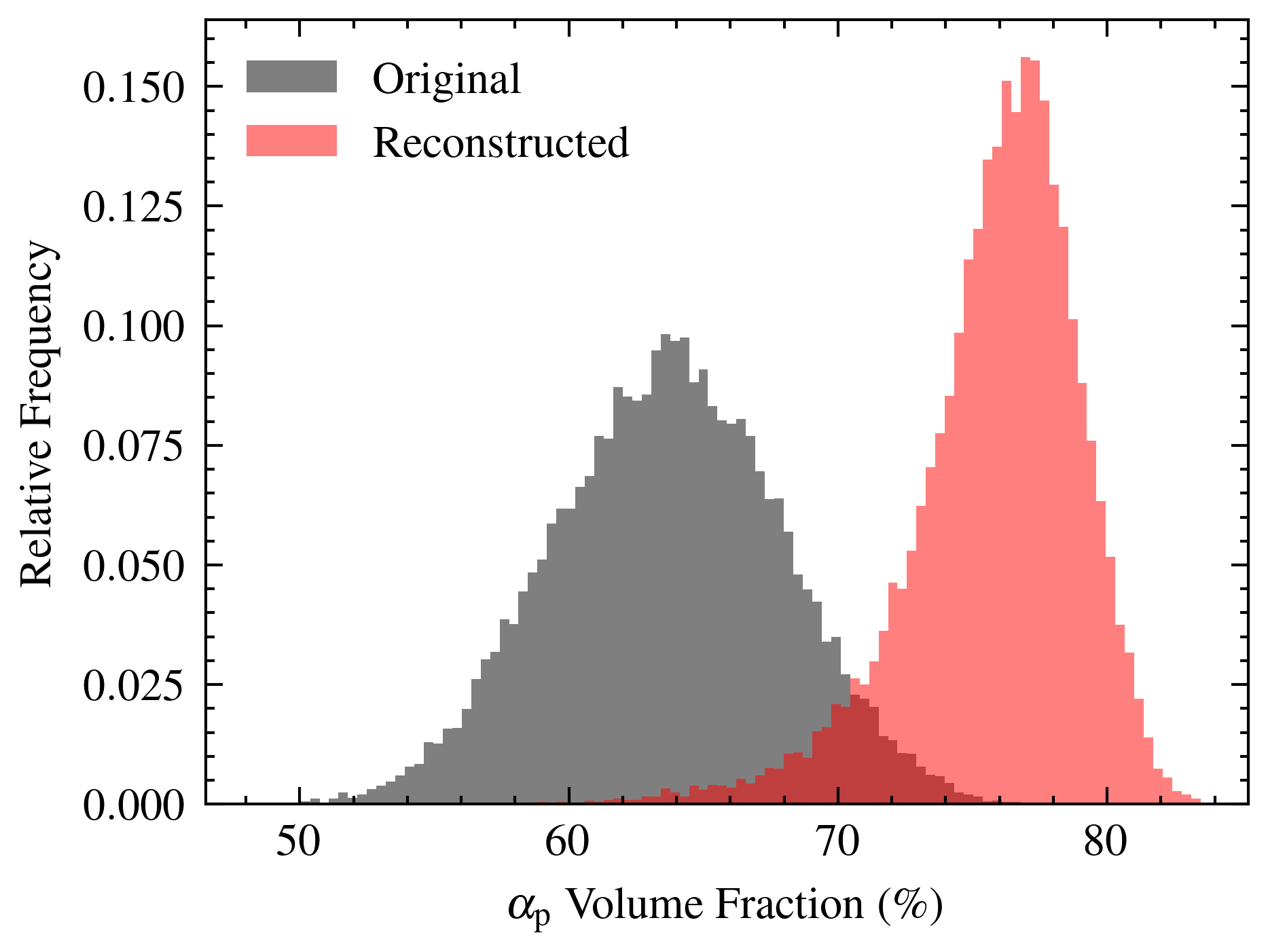}
        \label{subfig:vol_frac_bimodal}
    \end{subfigure}%
    \hfill
    \begin{subfigure}{0.45\textwidth}
        \centering
        \includegraphics[width=\textwidth]{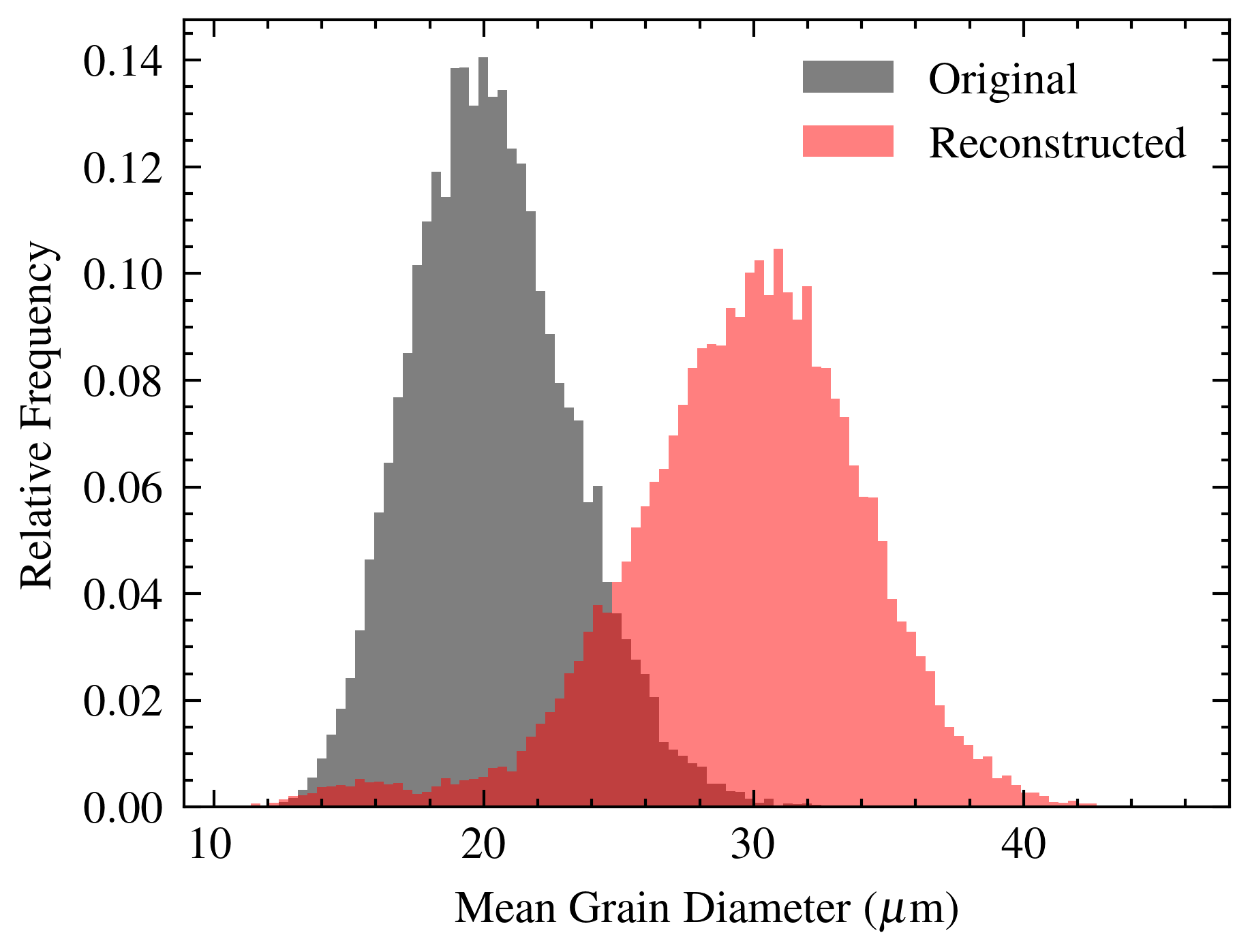}
        \label{subfig:grain_size_bimodal}
    \end{subfigure}
    \caption{Morphological metric distributions across the original image patches and their corresponding reconstructions for the bimodal dataset.}
    \label{fig:metrics_bimodal}
\end{figure}

Training the VAE on exclusively bimodal microstructures results in a similar smoothing effect as with the lamellar images, although this appears more pronounced in the bimodal case due to the nature of the fine scale features present in this case.
Figure~\ref{fig:resnet_recons_bimodal} shows some example reconstructions for the bimodal micrographs.
All $\upalpha_{\text{s}}$ laths retained in the prior $\upbeta$ grains are completely absent from the reconstructions and the outputs are effectively a mask for the $\upalpha_{\text{p}}$ grains.
This is confirmed to be a feature of the entire set of reconstructions in Figure~\ref{fig:metrics_bimodal}, which shows a drastic increase in both the $\upalpha_{\text{p}}$ volume fraction and mean $\upalpha_{\text{p}}$ grain size.

\subsection{Traversing the Encoded Space}
\label{subsec:traversal_results}

\begin{figure}[!ht]
    \centering
    \includegraphics[width=\textwidth]{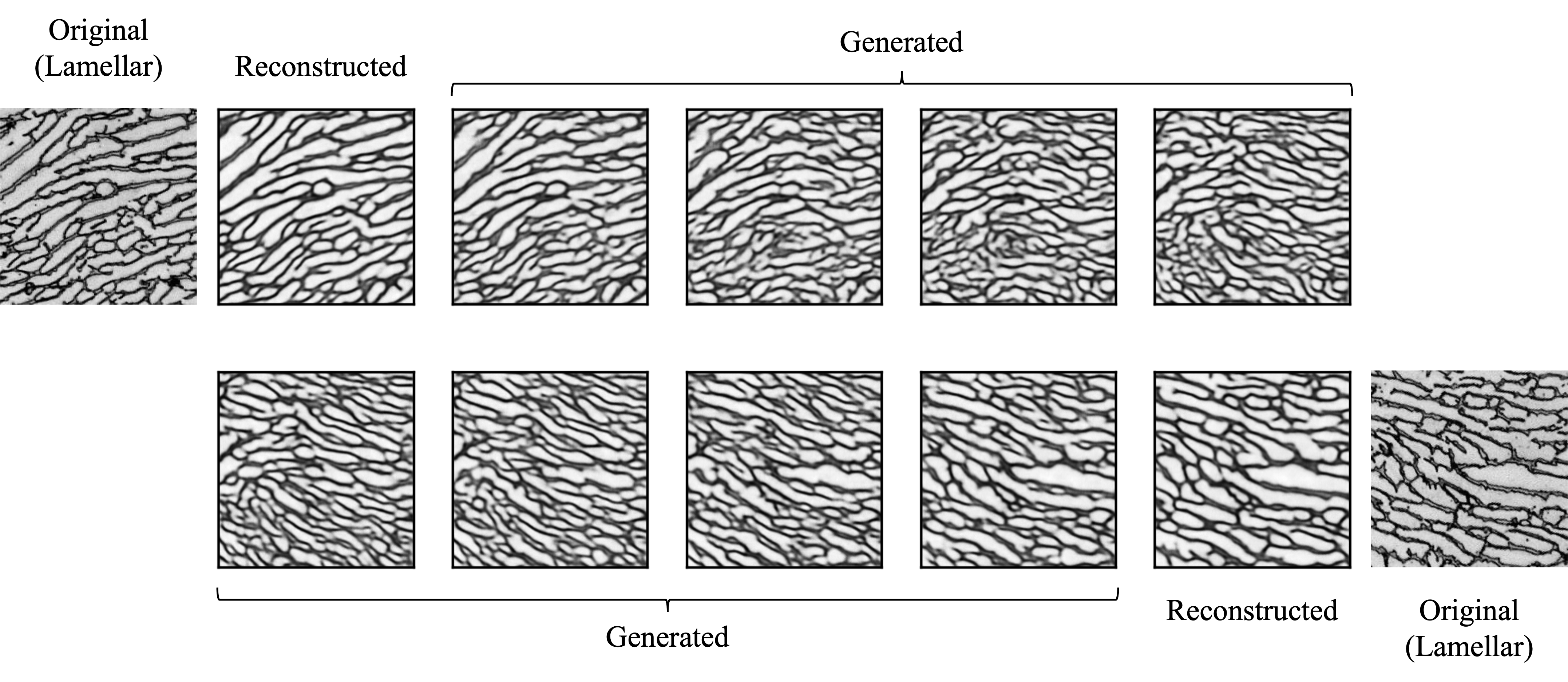}
    \caption{Reconstructions along linear path through the encoded space learned during training on exclusively lamellar image data from LFTi64BL.}
    \label{fig:lin_path_lfti64l}
\end{figure}

The encoded space was explored with the methods discussed in Section~\ref{subsec:traversing}.
Figure~\ref{fig:lin_path_lfti64l} shows reconstructions along a linear path between a pair of fingerprints from the training set.
A linear path between the fingerprints was constructed and 8 equispaced fingerprints were determined along the path.
These fingerprints were supplied to the trained decoder to generate the images in Figure~\ref{fig:lin_path_lfti64l}.

The fingerprints along this path show a smooth transition along the linear path, which is a direct result of the continuity of the latent space learned by the VAE\@.
These microstructures are synthetic and are not included in the training set, although they do have the same characteristics as the lamellar microstructures in the training set.

\begin{figure}[!ht]
    \centering
    \includegraphics[width=\textwidth]{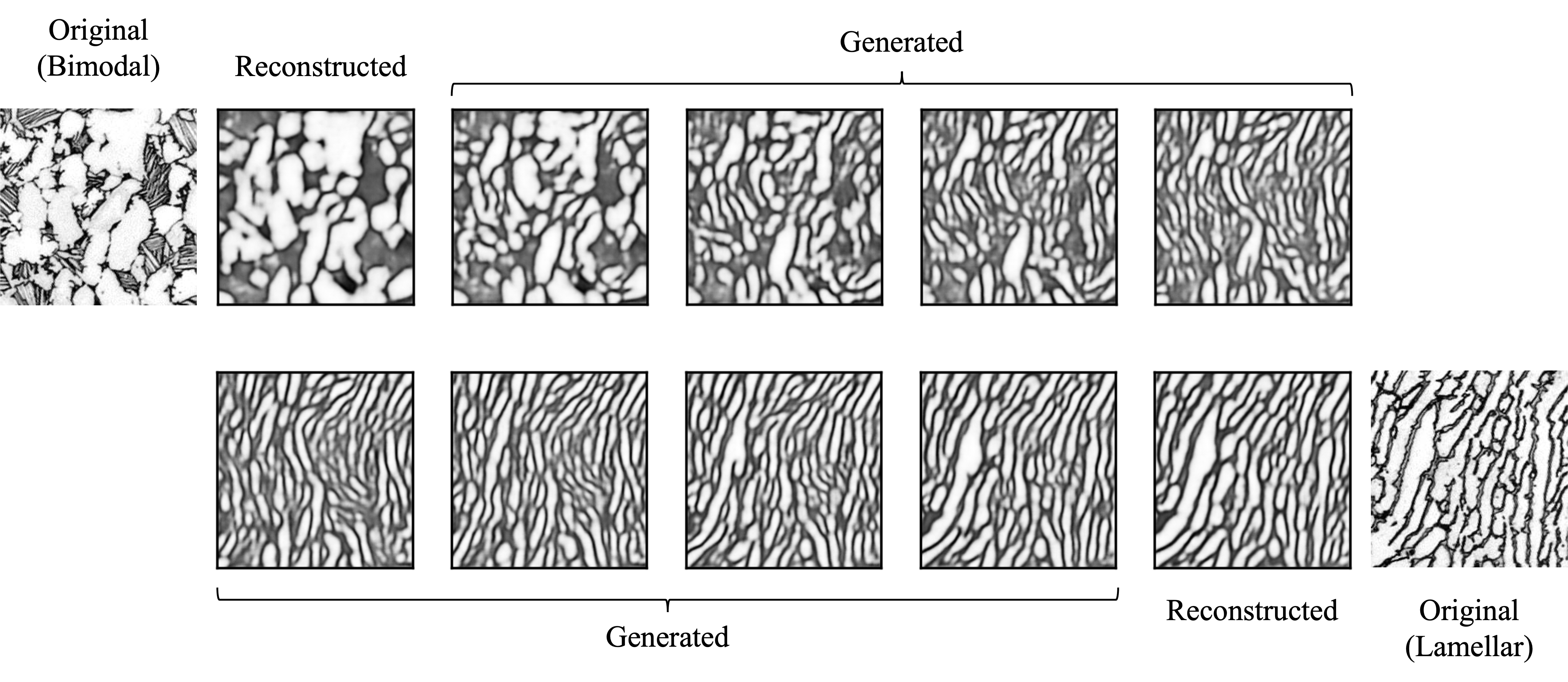}
    \caption{Reconstructions along linear path through the encoded space learned during training on the full LFTi64BL dataset.}
    \label{fig:lin_path_lfti64bl}
\end{figure}

Figure~\ref{fig:lin_path_lfti64bl} shows the output from the same method applied to the full LFTi64BL dataset, with the linear path defined between a lamellar and a bimodal microstructure.
The same smooth transition between the input microstructures is observed, however, intermediate microstructures along the path stray considerably away from the training set, particularly towards the centre of the path.
This is to be expected, but confirms that, for VAEs to generate convincing artificial micrographs, it is crucial that the training set be cohesive and not contain drastic variations in microstructure.
Otherwise, the latent space constructed is likely to contain microstructures that are not representative of the training set.

\begin{figure}[!ht]
    \centering
    \includegraphics[width=\textwidth]{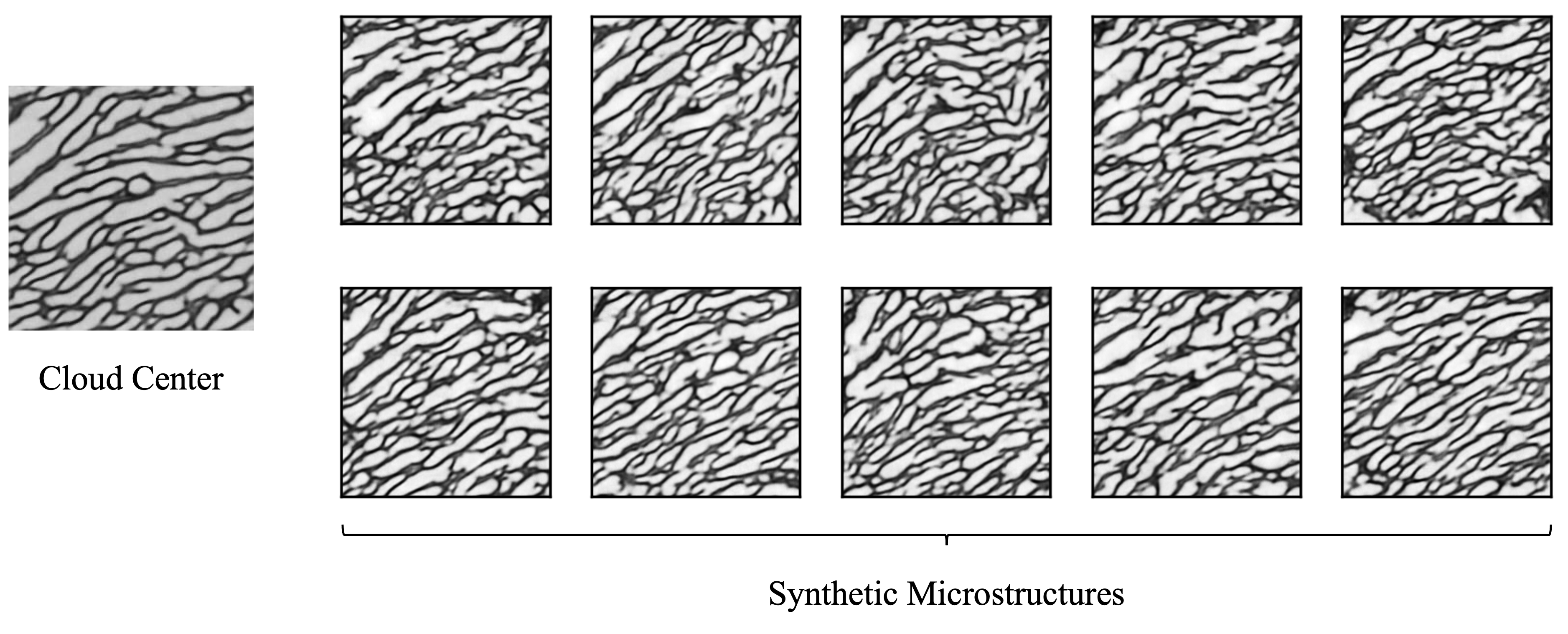}
    \caption{Synthetic microstructures generated from the addition of unit Gaussian noise, with a scale factor of $\gamma = 0.2$, to a known valid encoded representation, from which morphology is inherited.}
    \label{fig:rnd_noise}
\end{figure}

To visualise neighbouring microstructures localised around an individual fingerprint, a sample microstructure was randomly selected and supplied to the trained encoder.
The fingerprint obtained was perturbed with a small amount of Gaussian noise, as described in Section~\ref{subsec:traversing} with $\gamma = 0.2$, and provided to the trained decoder to generate a synthetic microstructure.
This process was repeated and Figure~\ref{fig:rnd_noise} shows 10 realisations of microstructures produced.

\begin{figure}[!ht]
    \centering
    \begin{subfigure}{0.14\textwidth}
        \centering
        \includegraphics[width=\textwidth]{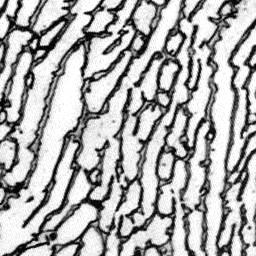}
        \caption{Original}
    \end{subfigure}
    \hfill
    \begin{subfigure}{0.14\textwidth}
        \centering
        \includegraphics[width=\textwidth]{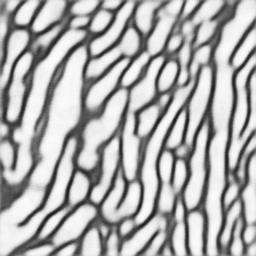}
        \caption{$\gamma = 0.05$}
    \end{subfigure}
    \hfill
    \begin{subfigure}{0.14\textwidth}
        \centering
        \includegraphics[width=\textwidth]{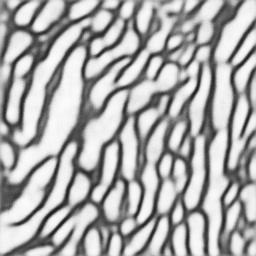}
        \caption{$\gamma = 0.1$}
    \end{subfigure}
    \hfill
    \begin{subfigure}{0.14\textwidth}
        \centering
        \includegraphics[width=\textwidth]{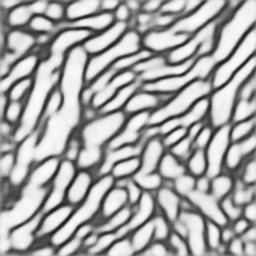}
        \caption{$\gamma = 0.2$}
    \end{subfigure}
    \hfill
    \begin{subfigure}{0.14\textwidth}
        \centering
        \includegraphics[width=\textwidth]{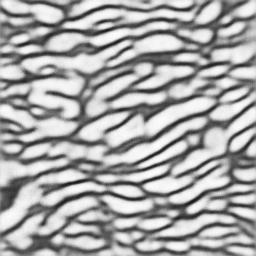}
        \caption{$\gamma = 0.5$}
    \end{subfigure}
    \hfill
    \begin{subfigure}{0.14\textwidth}
        \centering
        \includegraphics[width=\textwidth]{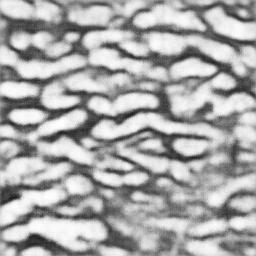}
        \caption{$\gamma = 1$}
    \end{subfigure}
    \caption{Synthetic micrograph examples generated via the Gaussian cloud method with various $\gamma$ values, compared with original micrograph.}
    \label{fig:rnd_gens}
\end{figure}

\begin{figure}[!ht]
    \centering
    \includegraphics[width=0.45\textwidth]{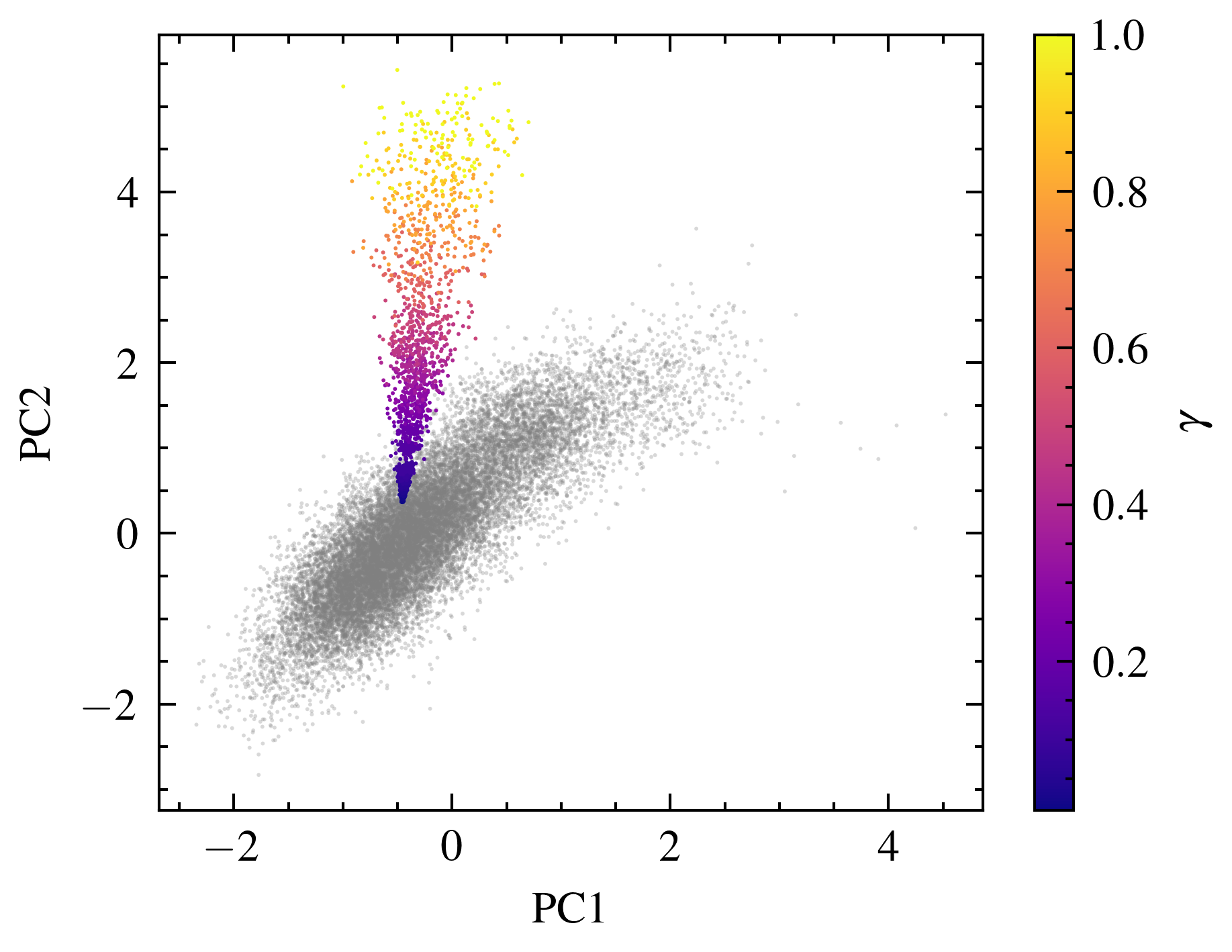}
    \caption{PCA applied to fingerprints leanred by the VAE alongside fingerprints constructed with the Gaussian cloud method. Original fingerprints are shown in grey and the colour map denotes the $\gamma$ values applied to generate the artifical fingerprints.}
    \label{fig:rnd_gen_pca}
\end{figure}

Each artificial microstructure possesses similar features to the input micrographs in terms of grain morphology and direction of lamellae with respect to the bottom edge of the image.
This shows that local fingerprints within the latent space are likely to possess similar microstructural features.
Figure~\ref{fig:rnd_gens} shows example micrographs constructed with increasing values of $\gamma$.
As $\gamma$ increases, images generated are no longer representative of the training dataset.
PCA was trained on the fingerprints constructed by the VAE and then used to transform the noise-perturbed fingerprints alongside the original fingerprints (see Figure~\ref{fig:rnd_gen_pca}).
This shows the noise-perturbed fingerprints emanating from a single point, which corresponds to the fingerprint to which the noise was applied, and we can see that noise added with a scale factor of $\gamma > 0.5$ are all completely outside the encoded space learned by the VAE\@.

\subsection{Metric Distributions Across the Encoded Space}
\label{subsec:tsne_metrics}

\begin{figure}[!ht]
    \centering
    \includegraphics[width=0.45\textwidth]{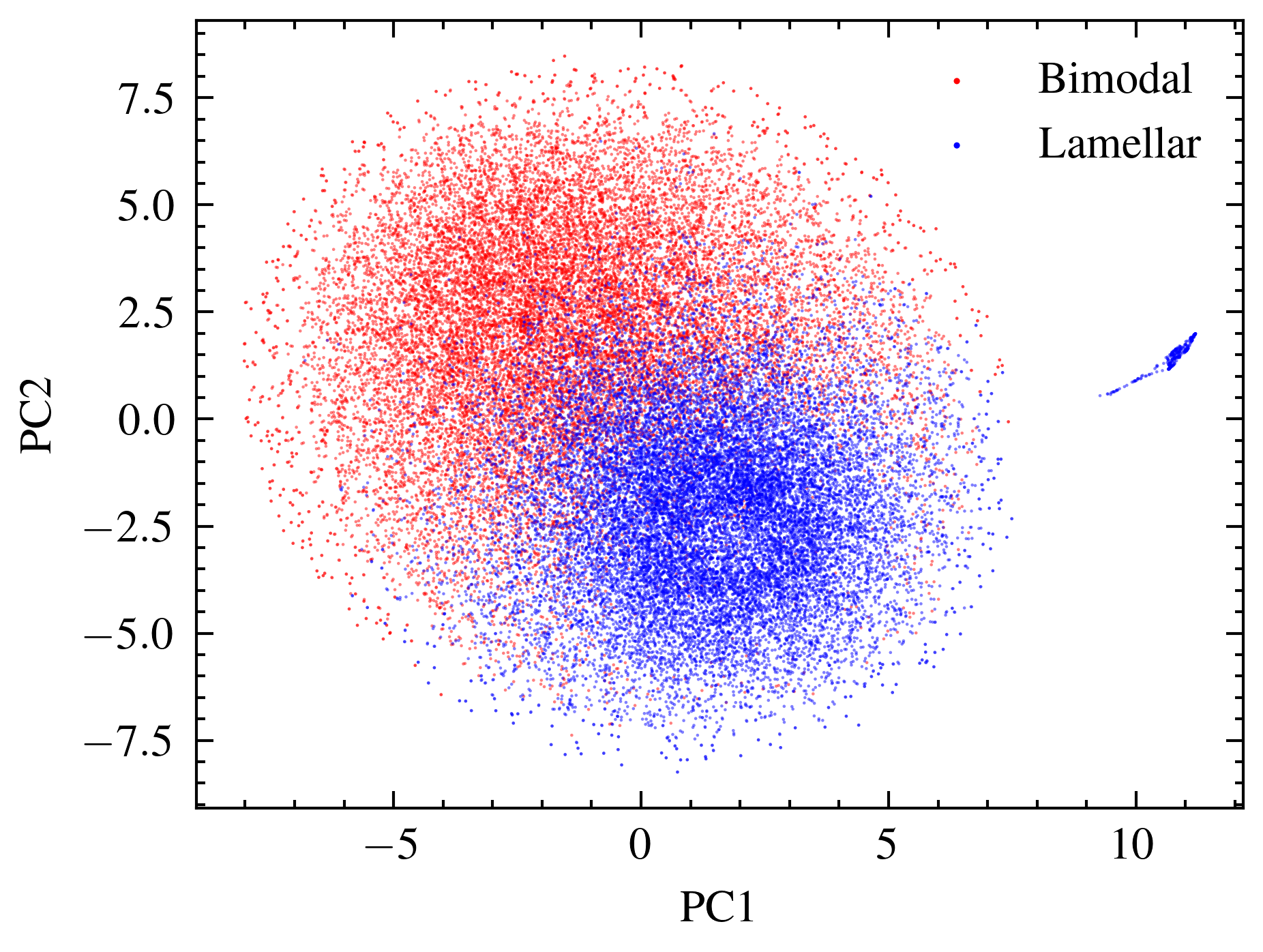}
    \caption{$t$-SNE with 2 components applied to encoded representations, where the colour map denotes classifications, with bimodal microstructures in red and lamellar microstructures in blue.}
    \label{fig:tsne_class_labels}
\end{figure}

Dimensionality reduction, in the form of $t$-SNE (described in Section~\ref{subsec:tsne}), was also applied to the original 256-dimensional fingerprints to reduce them down to 2-dimensional vectors.
This enables fingerprints to be plotted in a 2-dimensional scatter plot to visualise the entire latent space and distribution of metrics across the space.
Figure~\ref{fig:tsne_class_labels} shows the full LFTi64BL dataset reduced to 2-dimensions.
The colour map in this figure that illustrates the microstructure classification, with 0 denoting a bimodal microstructure and 1 denoting lamellar.
There is a strong clustering of each class, with only a small overlap between them, allowing fingerprints from each class to be easily separated.
Training an SVM with 90\% of the fingerprints allocated for training and 10\% reserved for testing yields a mean classification accuracy of 99.9\% $\pm$ 0.001 after 10-fold cross-validation.

\begin{figure}[!ht]
    \centering
    \begin{subfigure}{0.45\textwidth}
        \centering
        \includegraphics[width=\textwidth]{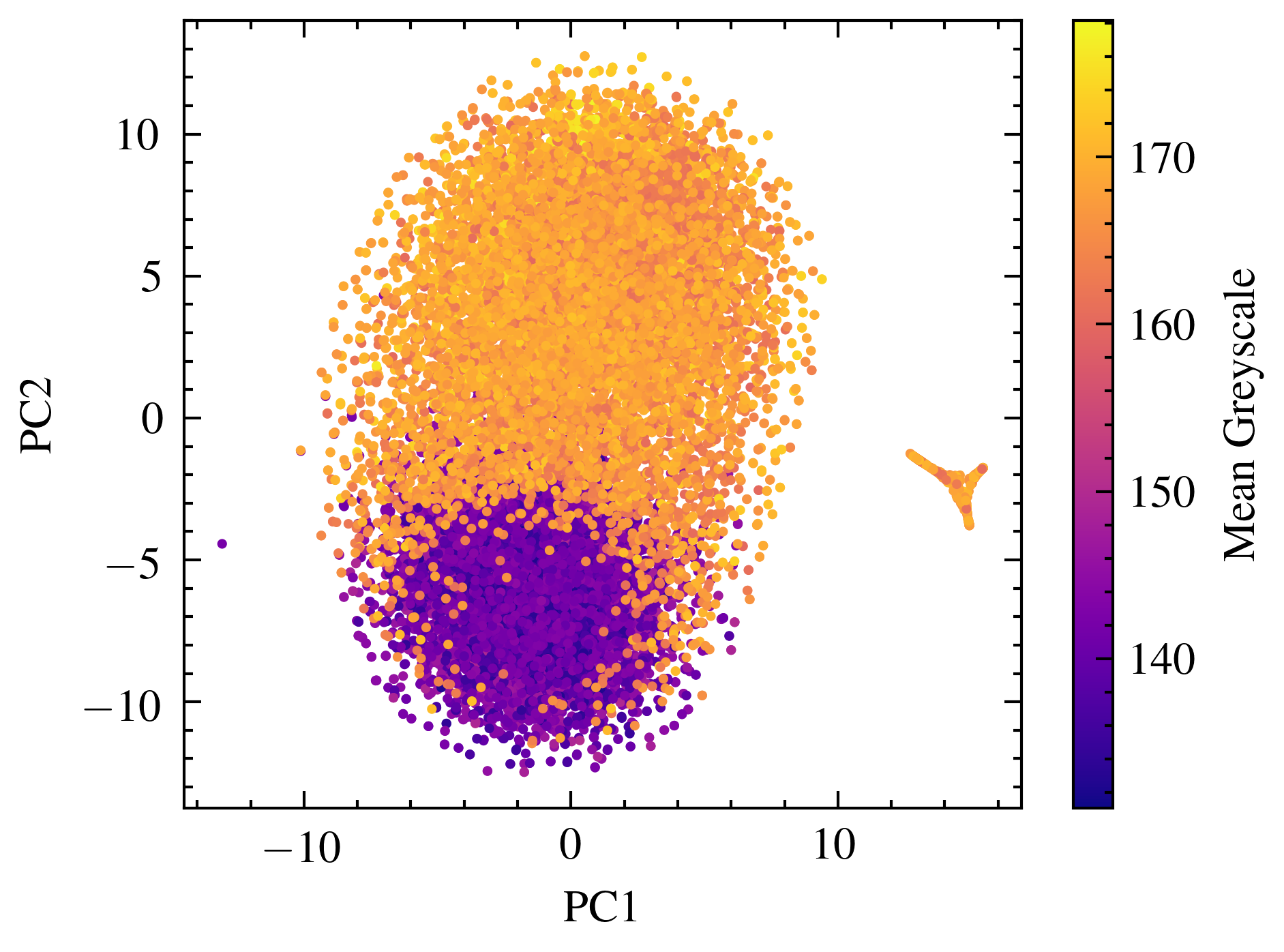}
        \caption{Lamellar greyscale prior to normalisation}
        \label{subfig:tsne_greyscale_lamellar_pre}
    \end{subfigure}%
    \hfill
    \begin{subfigure}{0.45\textwidth}
        \centering
        \includegraphics[width=\textwidth]{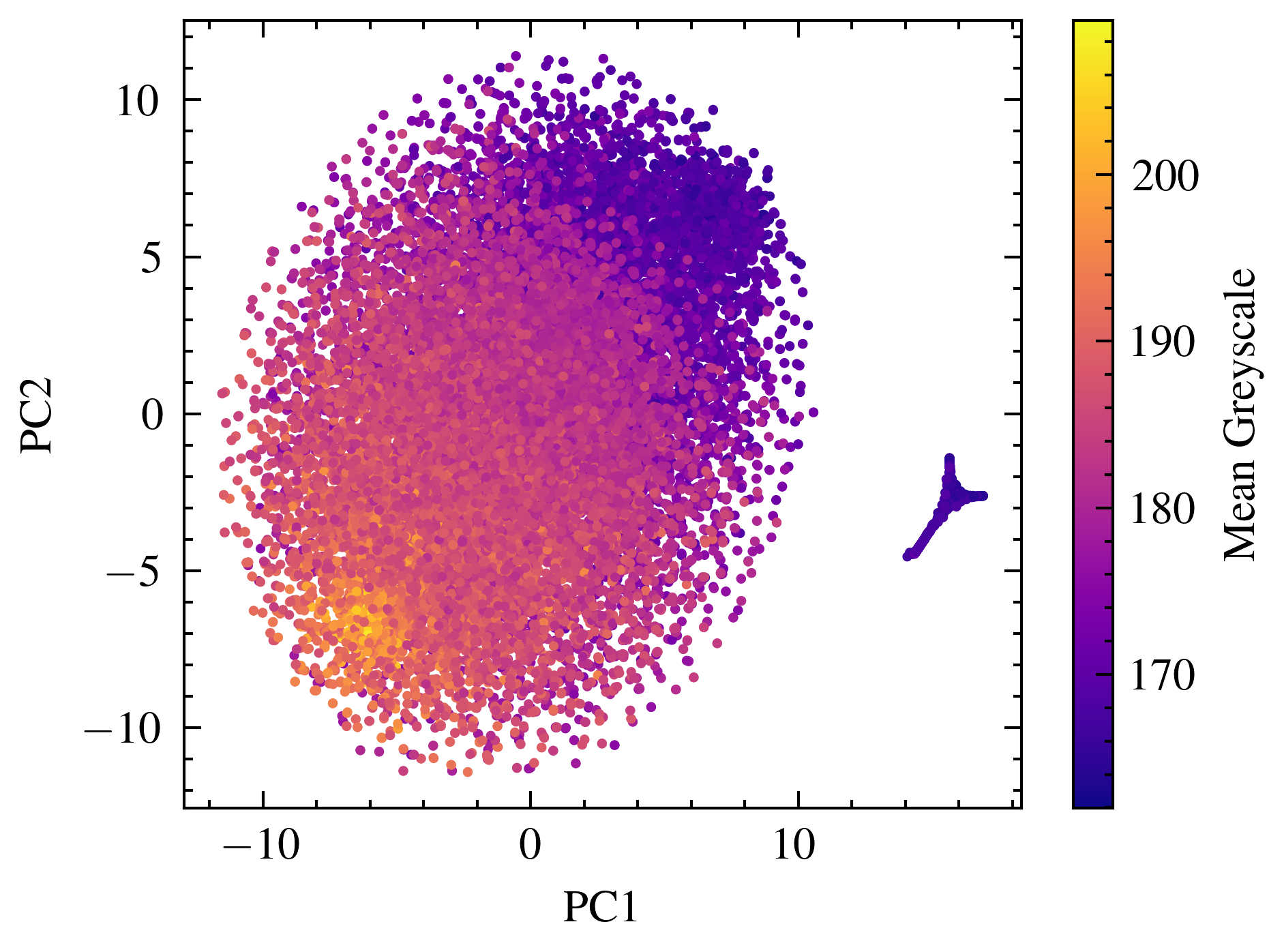}
        \caption{Lamellar greyscale after normalisation}
        \label{subfig:tsne_greyscale_lamellar_post}
    \end{subfigure}
    \centering
    \begin{subfigure}{0.45\textwidth}
        \centering
        \includegraphics[width=\textwidth]{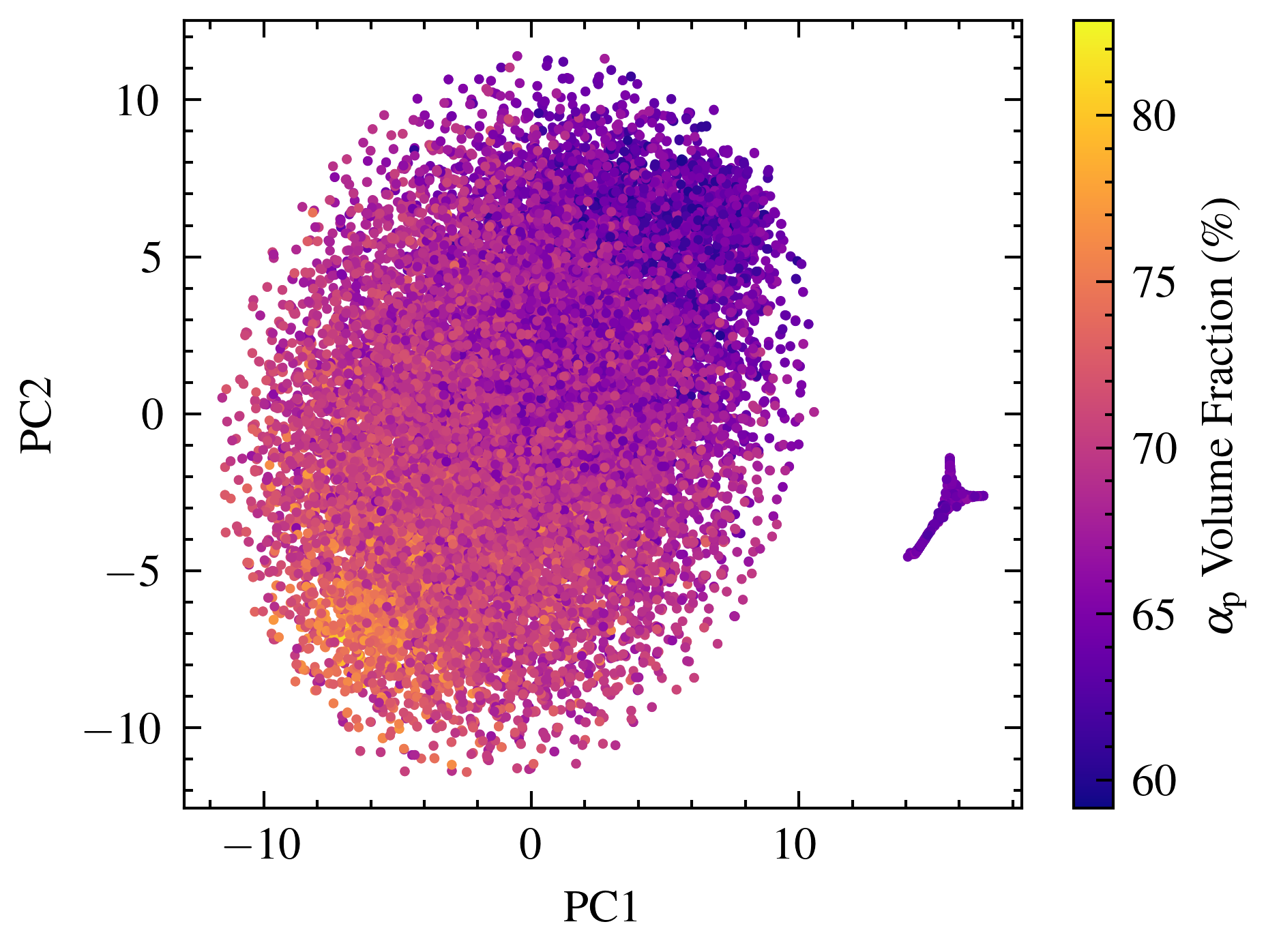}
        \caption{Lamellar $\upalpha_{\text{p}}$ volume fraction}
        \label{subfig:tsne_volfrac_lamellar}
    \end{subfigure}%
    \hfill
    \begin{subfigure}{0.45\textwidth}
        \centering
        \includegraphics[width=\textwidth]{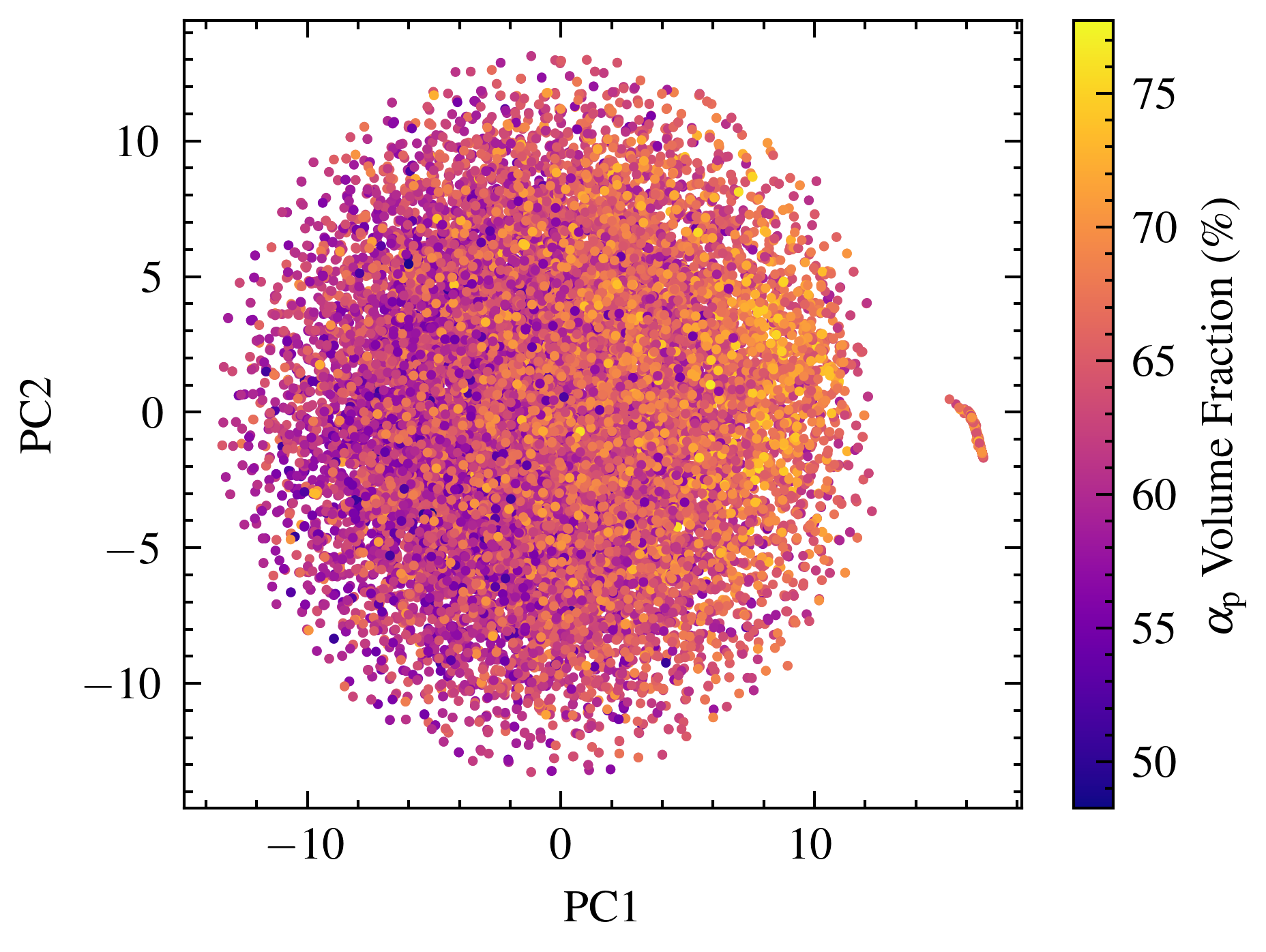}
        \caption{Bimodal $\upalpha_{\text{p}}$ volume fraction}
        \label{subfig:tsne_volfrac_bimodal}
    \end{subfigure}
    \begin{subfigure}{0.45\textwidth}
        \centering
        \includegraphics[width=\textwidth]{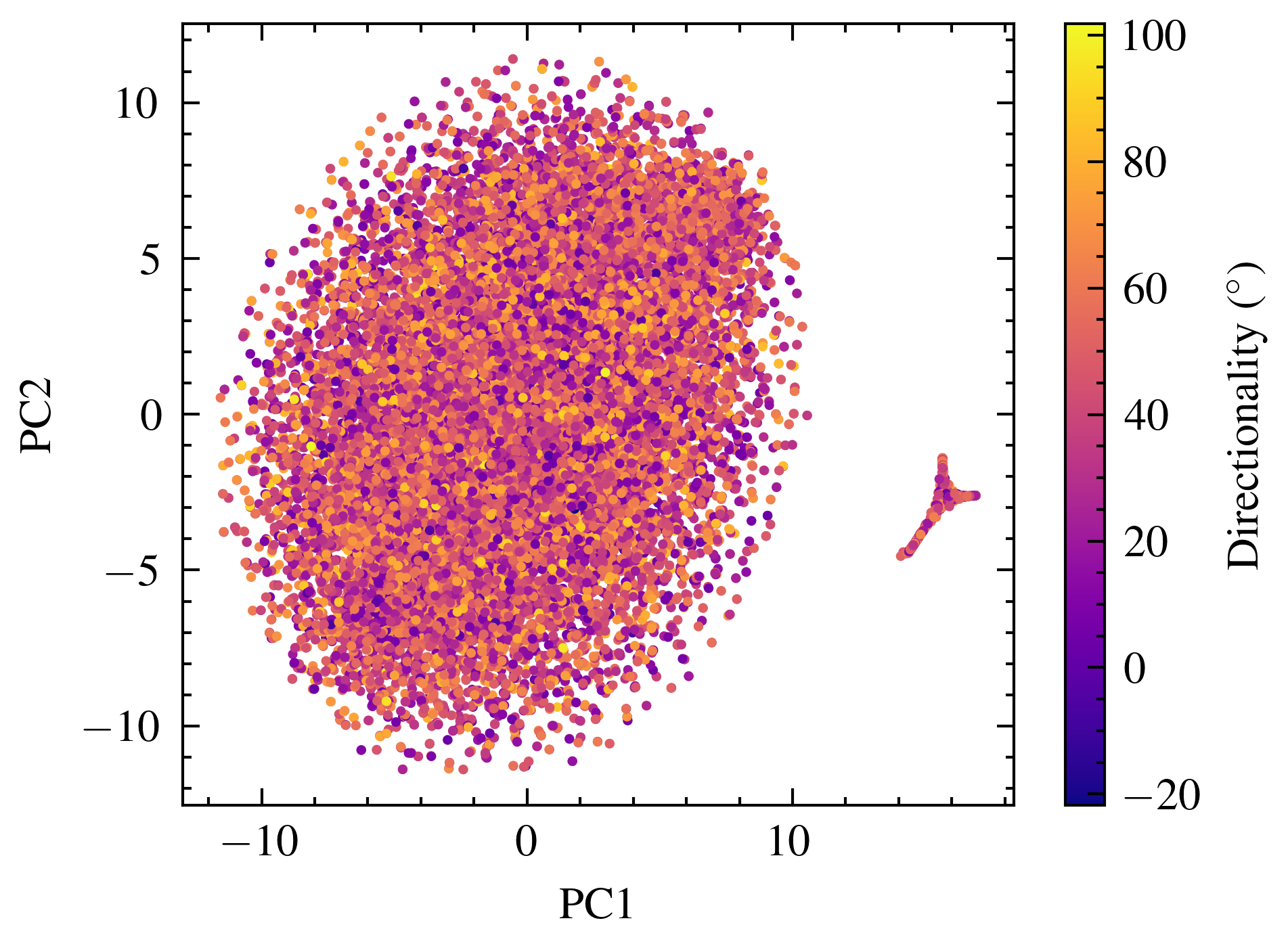}
        \caption{Lamellae directionality.}
        \label{subfig:tsne_direction}
    \end{subfigure}%
    \hfill
    \begin{subfigure}{0.45\textwidth}
        \centering
        \includegraphics[width=\textwidth]{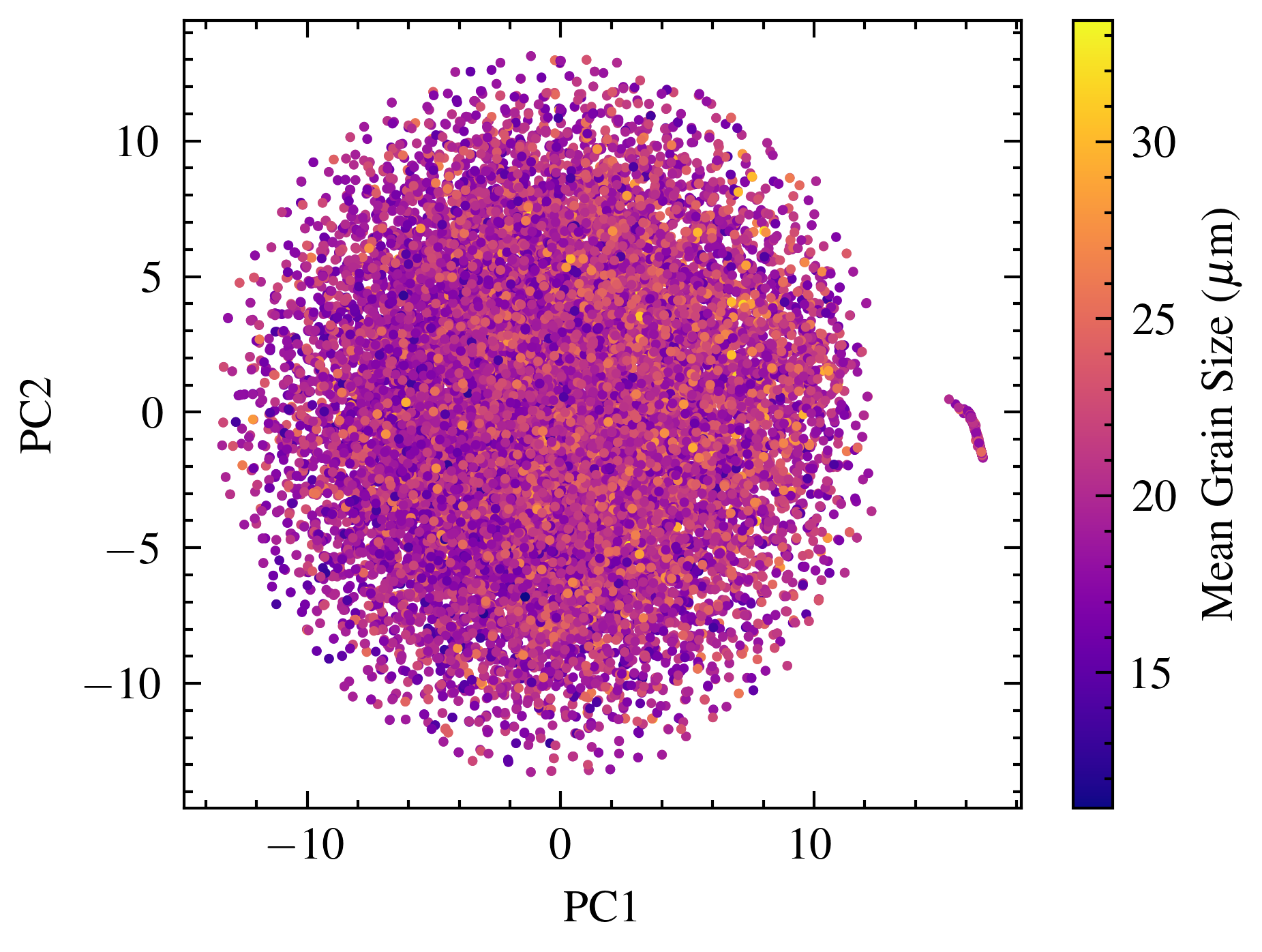}
        \caption{Bimodal mean $\upalpha_{\text{p}}$ grain size.}
        \label{subfig:tsne_grain_size}
    \end{subfigure}
    \caption{$t$-SNE with 2 components applied to encoded representations of the LFTi64BL dataset, with colour maps denoting various morphological metrics.}
    \label{fig:tsne_scatters}
\end{figure}

The same methodology for dimensionality reduction was applied to the lamellar and bimodal fingerprints separately to map distributions of various microstructural features.
Figures~\ref{subfig:tsne_greyscale_lamellar_pre} and~\ref{subfig:tsne_greyscale_lamellar_post} shows greyscale intensity from the lamellar micrographs, before and after normalisation.

Prior to normalisation, there is a strong clustering between two groups of images, heavily influenced by illumination during image capture.
After normalisation, greyscale intensity is more closely linked to volume fraction and is normally distributed across the dataset.
Dimensionality reduction then shows a smooth gradient of greyscale intensity across the latent space.
The plots with volume fraction yield similar results, although less pronounced.
Figures~\ref{subfig:tsne_volfrac_lamellar} and~\ref{subfig:tsne_volfrac_bimodal} shows plots of the encoded spaces for bimodal and lamellar microstructures, trained separately.

Finally, we look at directionality (discussed in Section~\ref{subsubsec:direction}) for lamellar microstructures and grain size (Section~\ref{subsubsec:grain_size}) for bimodal microstructures as morphological features of interest.
Figure~\ref{subfig:tsne_direction} shows the $t$-SNE plot with a colour map corresponding to lamellae direction.

This appears to show random scatter across the encoded space, in contrast to the Gaussian cloud for image generation that seems to reconstruct images with similar direction, when perturbing fingerprints with a small amount of Gaussian noise.
Nearest neighbours are shown to have similar directionality, but this is contained within small regions of the latent space and not universal.
This random distribution of directionality may be useful in practice, though, as this implies that when encoding the microstructural information in such a manner, directionality can be effectively ignored and sample orientation when imaging would not be of any concern.

Figure~\ref{subfig:tsne_grain_size} shows the $t$-SNE plot for the bimodal fingerprints with a colour map denoting grain size.
Here, we see a gradient of grain size across the latent space, suggesting that nearest neighbours will share similar morphologies.

In each $t$-SNE reduction, there is a small group of fingerprints that are separated away from the main cluster.
There are no discernible differences between the microstructures that result in these fingerprints and those that form the main cluster.
Despite these microstructures being included in the training set, their reconstructions after training are effectively just noise.
This is a caveat of the VAE training process and not the dimensionality reduction.
Removing these outliers from the set of fingerprints and rerunning the $t$-SNE with the remaining fingerprints results in retention of the main cluster and a complete absence of the secondary cluster.
Removing the images that reside in the secondary cluster from the training set and retraining the VAE from scratch results in a new cluster forming, with a different morphology, that contains a different set of images.
Nonetheless, we are able to explore the encoded space of valid reconstructions, using the $t$-SNE as a guide to identify suitably encoded microstructures.

\subsection{Metric Predictions from Encoded Representations}
\label{subsec:svr_pred}

If an approach to microstructural fingerprinting is successfully encoding the essence of the microstructure, then we might reasonably expect to be able to recover key features of the microstructure, such as grain size, from the encoded fingerprint.
A strong correlation between the fingerprint and a given property also opens the possibility of reversing the inferential process and asking what fingerprints (or which regions of latent space) would correspond to a microstructure exhibiting a given property of interest (with the possibility with a VAE of then reconstructing an image of the corresponding microstructure).

\begin{figure}[!ht]
    \centering
    \begin{subfigure}{0.45\textwidth}
        \centering
        \includegraphics[width=\textwidth]{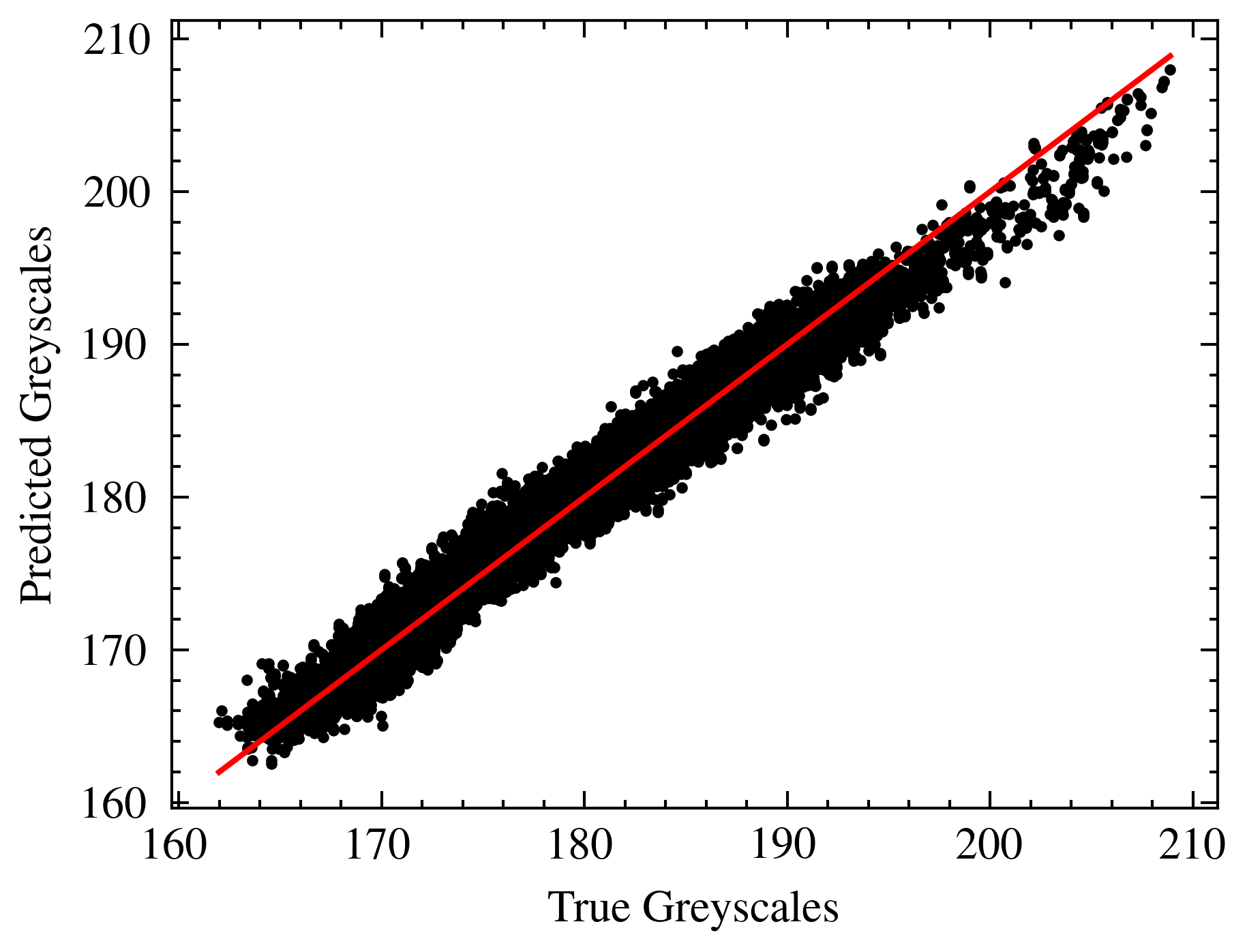}
        \caption{Lamellar greyscale intensity}
    \end{subfigure}%
    \hfill
    \begin{subfigure}{0.45\textwidth}
        \centering
        \includegraphics[width=\textwidth]{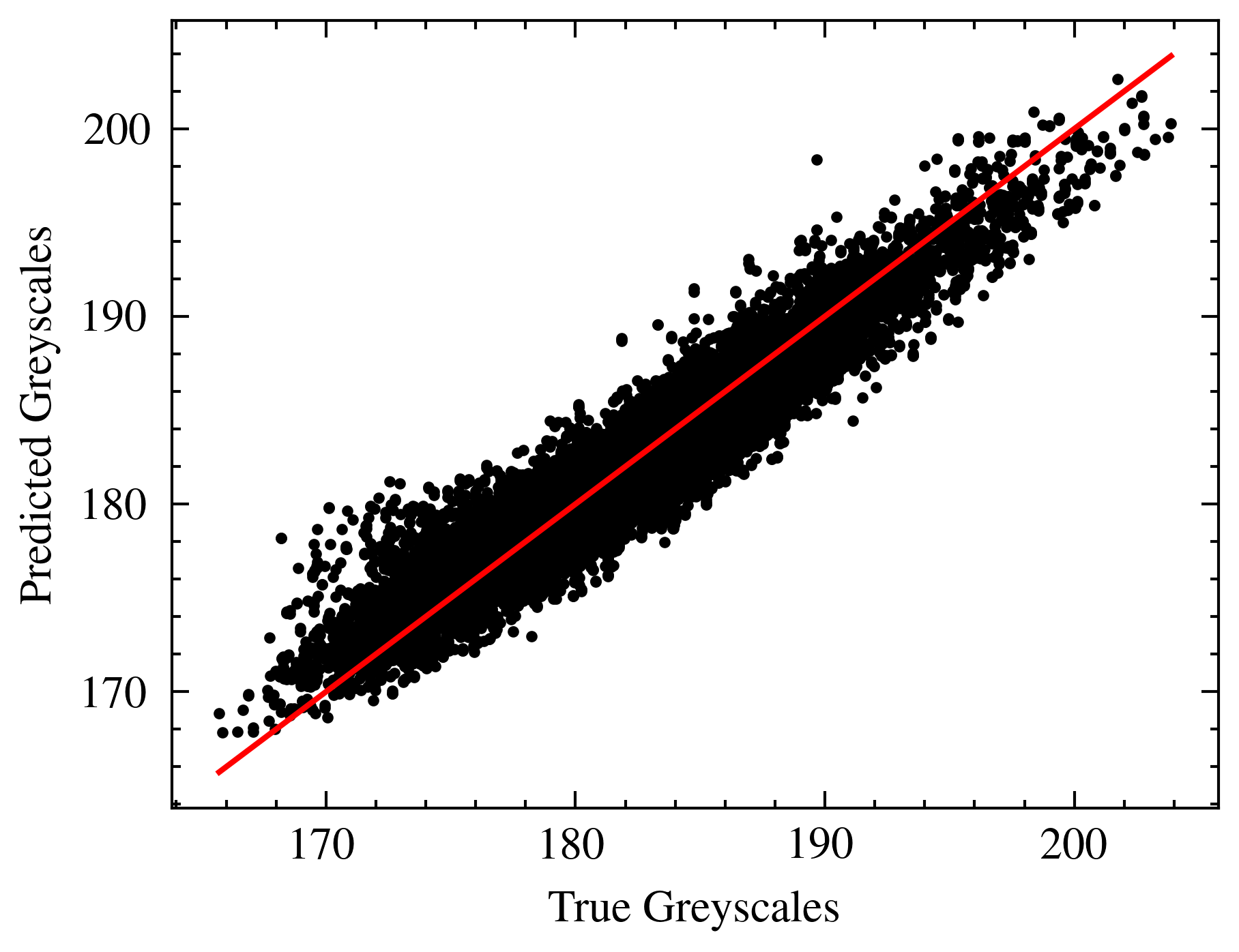}
        \caption{Bimodal greyscale intensity}
    \end{subfigure}
    \begin{subfigure}{0.45\textwidth}
        \centering
        \includegraphics[width=\textwidth]{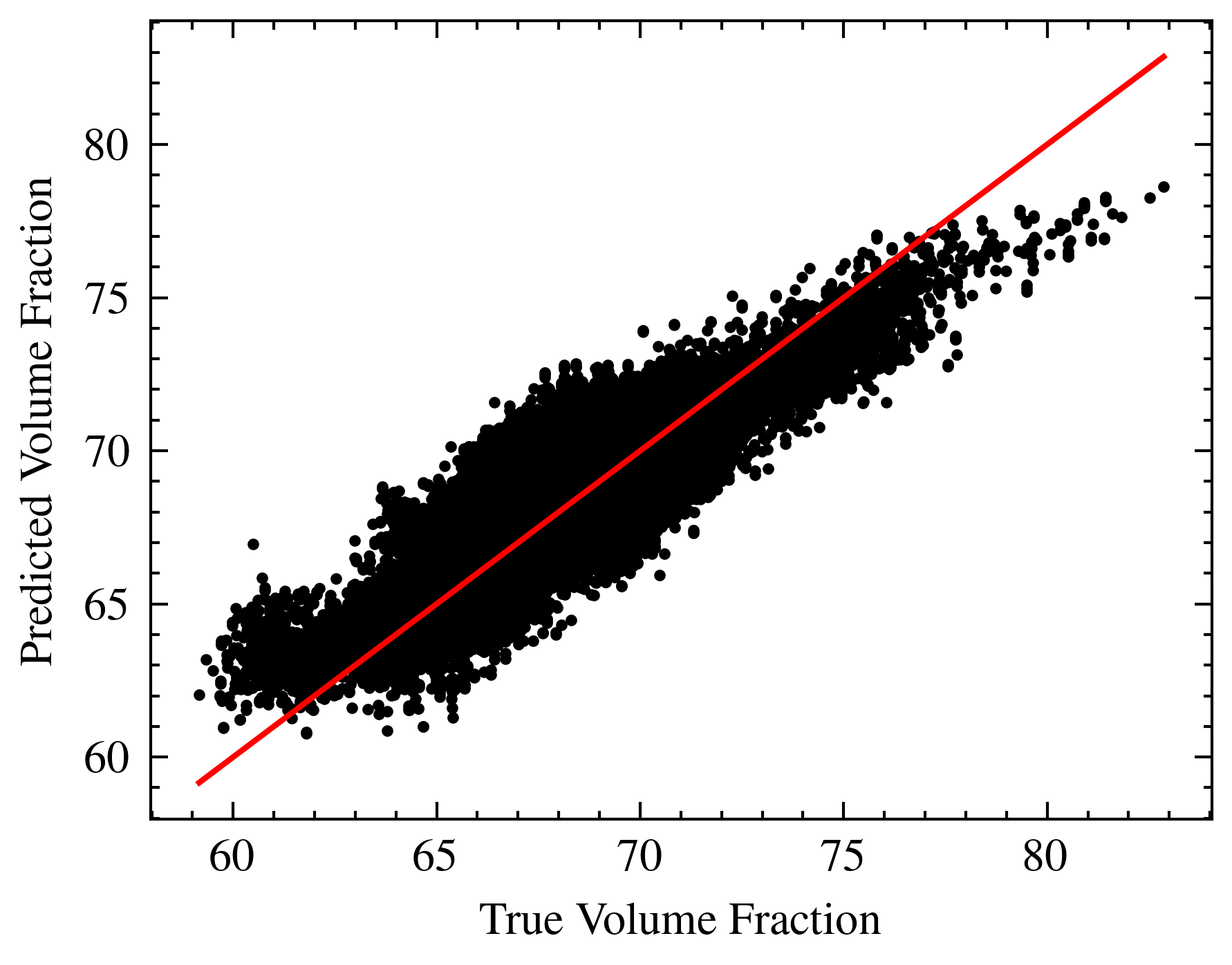}
        \caption{Lamellar volume fraction $\upalpha_{\text{p}}$}
    \end{subfigure}%
    \hfill
    \begin{subfigure}{0.45\textwidth}
        \centering
        \includegraphics[width=\textwidth]{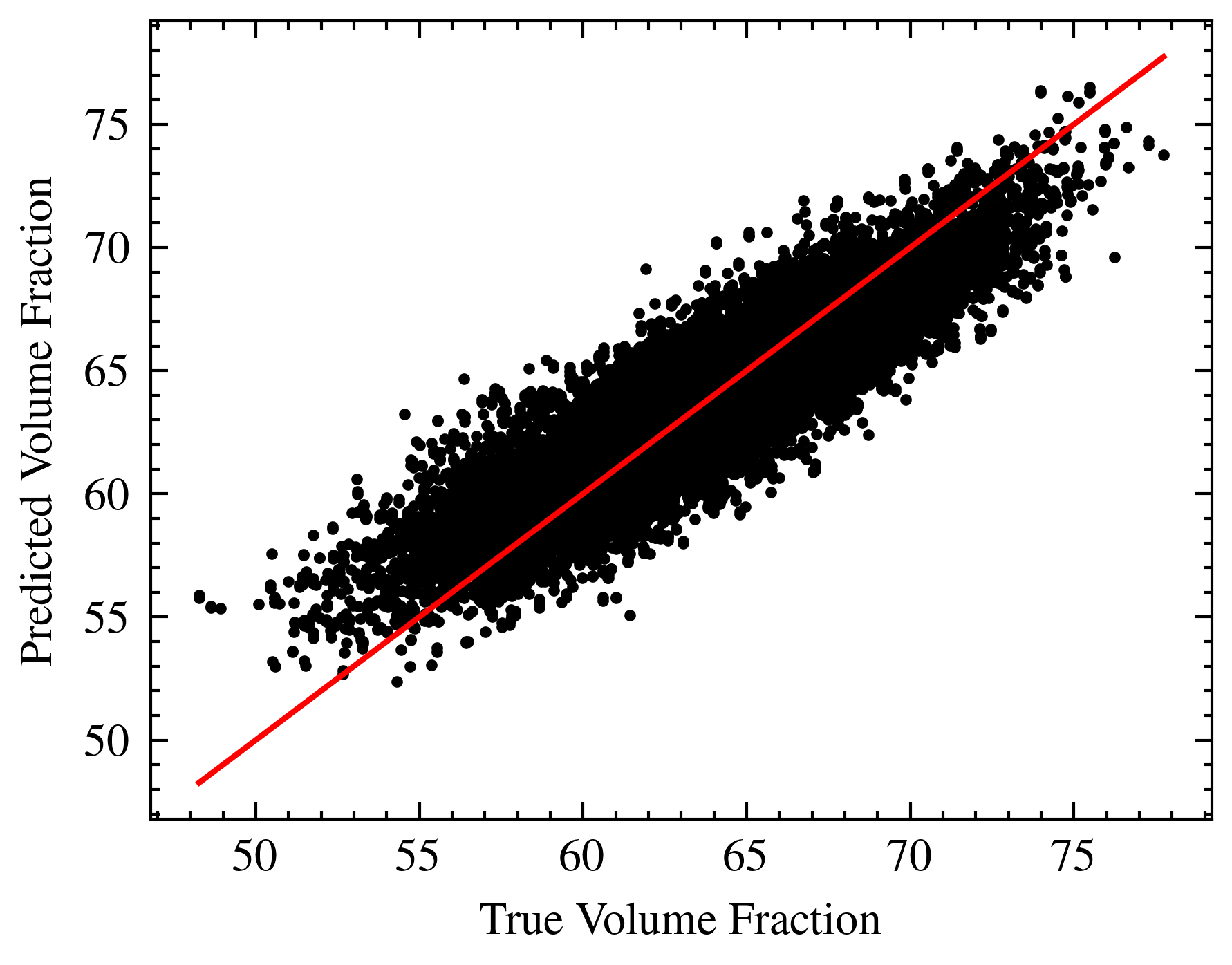}
        \caption{Bimodal volume fraction $\upalpha_{\text{p}}$}
    \end{subfigure}
    \begin{subfigure}{0.45\textwidth}
        \centering
        \includegraphics[width=\textwidth]{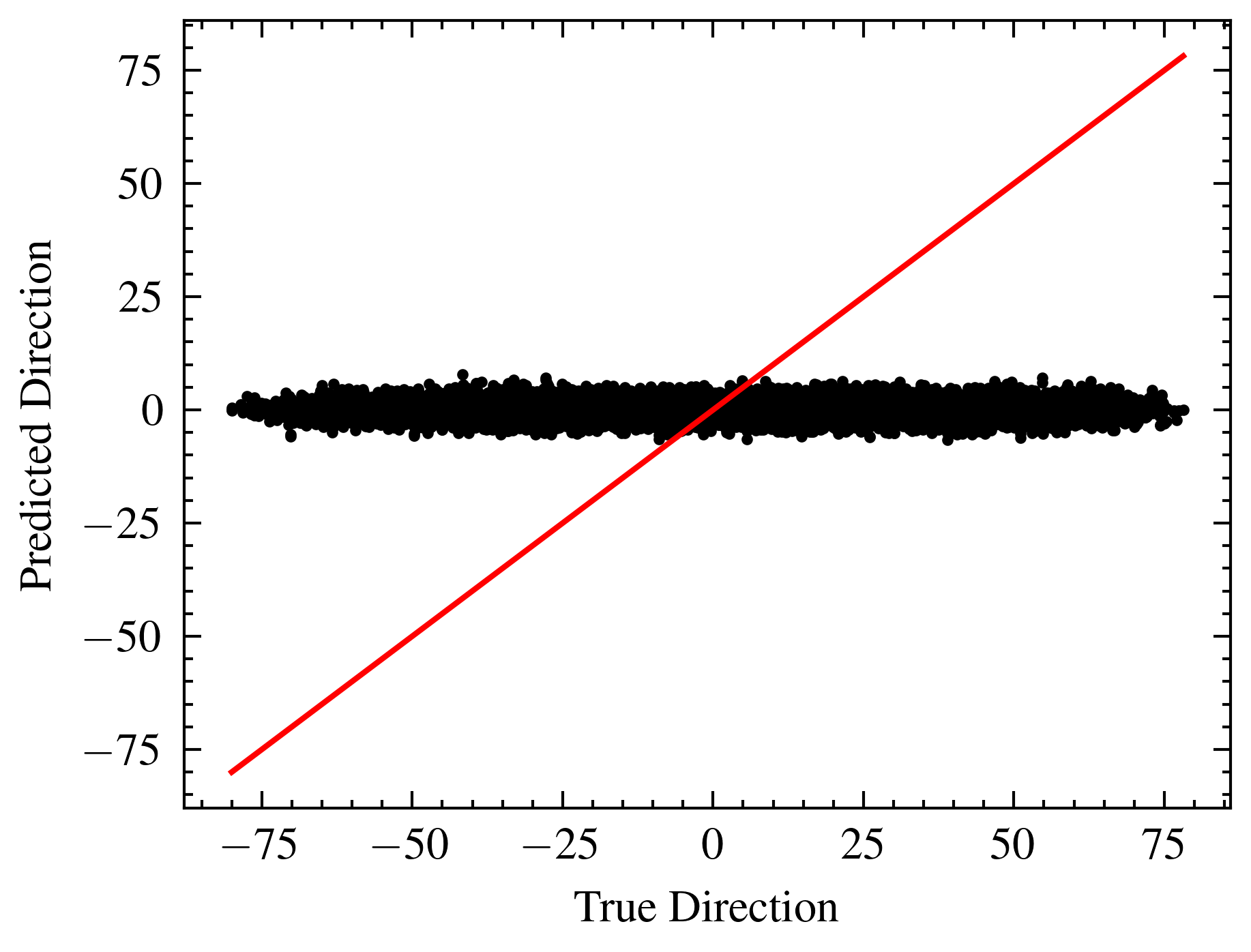}
        \caption{Lamellae direction}
    \end{subfigure}%
    \hfill
    \begin{subfigure}{0.45\textwidth}
        \centering
        \includegraphics[width=\textwidth]{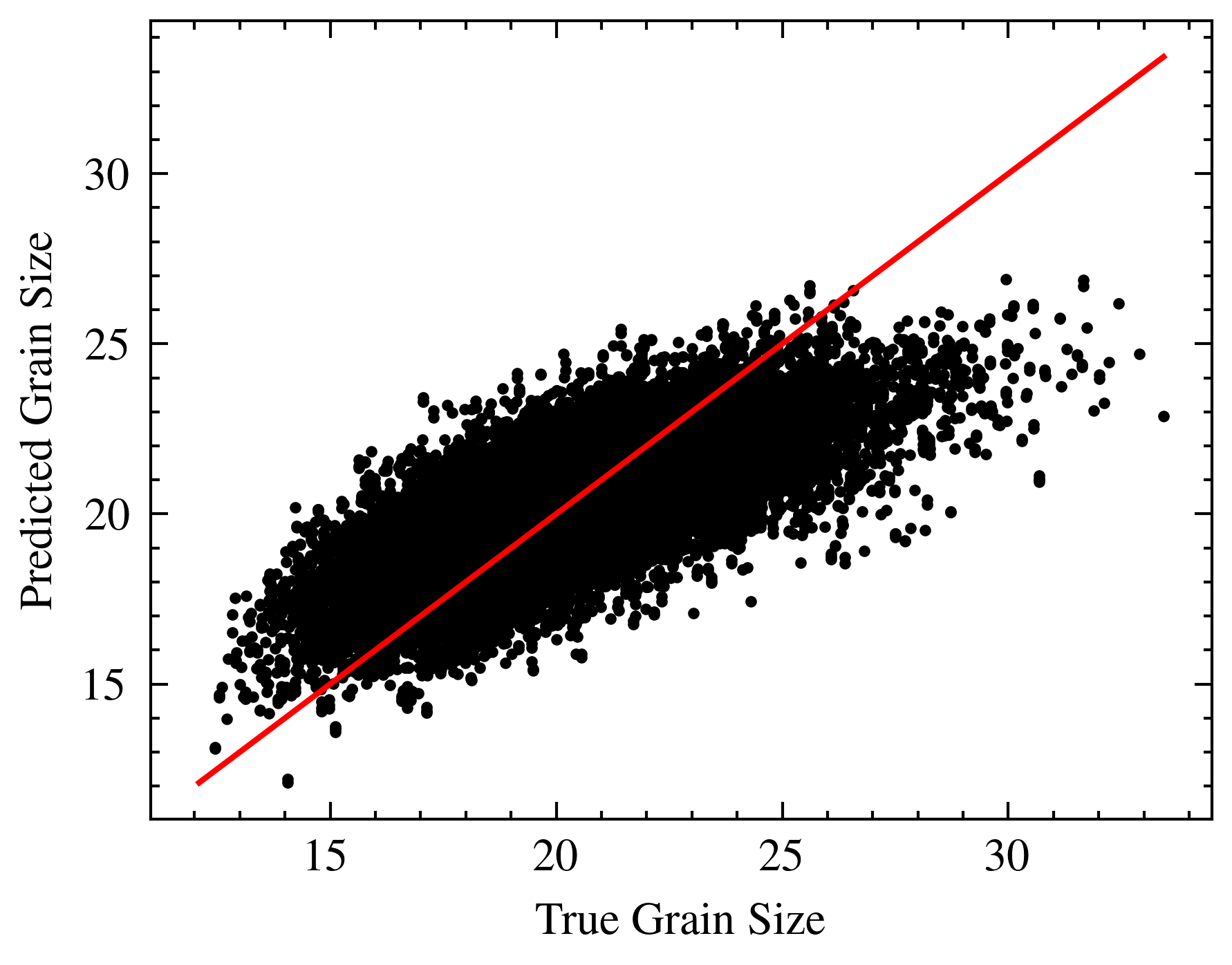}
        \caption{Bimodal grain size}
    \end{subfigure}
    \caption{Scatter plots to illustrate combined 10-fold cross validation SVR predictions of various morphological metrics with the line $y=x$ added to highlight where true predictions should lie.}
    \label{fig:svr_linear}
\end{figure}

The predictability of morphological metrics across the encoded space was validated with Support Vector Regression (SVR), described in Section~\ref{subsec:svr}.
The fingerprints output from the RestNet18 VAE are randomly split into 90\% training data and 10\% test data.
This is repeated 10 times to perform 10-fold cross-validation.
The SVR is then trained on the training set and outputs for the test set are compared with the true values to measure the prediction accuracy.
Figure~\ref{fig:svr_linear} shows scatter plots for the SVR predictions across the 10 train-test splits combined, against the true values, with the line $y=x$ plotted in red to illustrate deviation from exact predictions.
Table~\ref{tab:svr_error} then shows percentage error and standard deviation of the predictions relative to the true values.

\begin{table}[!ht]
    \centering
    \begin{tabular}{ |c|c|c| }
        \hline
        \textbf{Class}                  & \textbf{Metric}                       & \textbf{Mean Percentage Error} \\
        \hline
        \multirow{3}{*}{Lamellar}       & Greyscale Intensity                   & 0.60 \% $\pm$ 0.01           \\
        \cline{2-3}
        & Volume Fraction $\upalpha_{\text{p}}$ (\%)   & 1.83 \% $\pm$ 0.03           \\
        \cline{2-3}
        & Directionality ($^\circ$)                        & 148 \% $\pm$ 29          \\
        \hline
        \multirow{3}{*}{Bimodal}       & Greyscale Intensity                   & 0.75 \% $\pm$ 0.01           \\
        \cline{2-3}
        & Volume Fraction $\upalpha_{\text{p}}$ (\%)   & 2.50 \% $\pm$ 0.03           \\
        \cline{2-3}
        & Grain Size ($\mu$m)                   & 7.91 \% $\pm$ 0.12           \\
        \hline
    \end{tabular}
    \caption{Quantitative analysis of error for SVR predictions.}
    \label{tab:svr_error}
\end{table}


\section{Discussion}
\label{sec:discussion}

Reconstructions from the proposed ResNet18 VAE were shown to accurately identify grain boundaries, albeit with some smoothing.
This information loss is acceptable for correlating with the morphological features discussed, as these features are still captured.
However, prediction of properties that rely upon small scale features, such as fine $\upalpha_{\text{s}}$ laths, may become hindered.
This could be alleviated with higher resolution images and patches with a higher magnification.

The encoded space learned by the VAE is continuous and contains valid microstructural fingerprints that are not included in the training set.
Two methods were used to explore the encoded space as a tool for generating artificial microstructures.
The linear path method illustrates the continuity of the space and shows a smooth transition between microstructures, when linearly interpolating between two known fingerprints.
With a training set containing only lamellar microstructures, reconstructions along this path are completely artificial but still resemble lamellar microstructures.
Once the training set is expanded to include both bimodal and lamellar microstructures, the linear interpolation between a bimodal fingerprint and a lamellar fingerprint yields a blend between the two morphologies.
These can still be perceived as valid microstructures, but it is important to note that these are not representative of any data in the training set and VAEs trained in such a way should be interpolated carefully.

Gaussian noise was also used to explore the encoded space more locally around an individual fingerprint.
Iteratively perturbing a learned fingerprint and supplying the output to the trained decoder yields another set of artificially generated microstructures.
In this case, the decoder generates microstructures with similar features to the input image, including the lamellae thickness and direction relative to the bottom of the image, provided that a suitably small amount of noise is applied.
This behaviour is only observed locally for some features, such as lamellae direction, which is randomly distributed across the encoded space.
This may be due to the fact that the VAE was trained on patches with random rotations applied, making the VAE rotationally invariant in this case.
This removes any bias towards sample orientation during image capture.
Adding large amounts of noise would result in variations of such features.

Dimensionality reduction via $t$-SNE was performed to reduce the fingerprints to 2 dimensions.
This enables them to be plotted in a 2-dimensional scatter plot to visualise the distribution of morphological features across the encoded space.
The fingerprint position in the encoded space appears to dependent significantly on the volume fraction of $\upalpha_{\text{p}}$, relative to other morphological metrics.
This was confirmed by SVR predictions, which was trained to predict volume fraction with an average percentage error of 1.83 \% $\pm$ 0.03 for the lamellar dataset and 2.50 \% $\pm$ 0.03 for the bimodal dataset.
Grain size was also highly correlated with the encoded representations and SVR was able to predict grain size with reasonable accuracy, giving an average percentage error of 7.91 \% $\pm$ 0.12.
Directionality of lamellae appeared to be distributed randomly across the encoded space from the $t$-SNE plots and this was also confirmed with SVR, which was unable to learn any trends in the data and repeatedly predicted the mean direction for each fingerprint, resulting in an average percentage error of 148 \% $\pm$ 29.
In practice, the random nature of the distribution of lamellae direction could be a useful feature, as this implies that sample orientation under the microscope can be effectively ignored when capturing a dataset for VAE without any bias being introduced.


\section{Conclusions}
\label{sec:conclusions}

The latent space of microstructural fingerprints generated from variational autoencoders (VAEs) has been explored to further our understanding of how VAEs encode feature information from microstructural image data.

\begin{itemize}
    \item We show that a variational autoencoder (VAE) architecture based on ResNet18 is able to produce accurate reconstructions of microstructures (up to some smoothing).
    \item Interpolation of fingerprints along a linear path in latent space and random perturbations about fingerprints of input microstructures resulted in the generation of plausible synthetic microstructures, demonstrating the suitability of the VAE for smooth representation of microstructure.
    \item Fingerprints constructed by the trained VAE are shown to correlate with morphological features, including $\upalpha{\text{p}}$ volume fraction and grain size, using support vector regression (SVR) with 10-fold cross-validation.
    \item Principal component analysis (PCA) is shown to provide useful insight into the amount of noise that can be added to known fingerprints before the perturbed fingerprints become disconnected from the learned latent space.
\end{itemize}

Our study based on a set of micrographs of titanium alloy microstructure demonstrates that a VAE can encode material microstructures to produce fingerprints that exhibit several of the key features required in a general purpose fingerprint.
Important next steps will be to test this approach on a broader class of microstructures and to apply it to datasets exhibiting variation in both microstructure and measured properties, to allow exploration of the suitability of VAEs for predicting microstructure-process-property relationships in an explainable framework.
We also note that generative adversarial networks (GANs) are found to generate high-fidelity synthetic microstructures, with strong statistical similarity to the training data, even in a limited data regime~\cite{gowtham2023}.
It would be interesting to consider synthetic images output from a GAN as inputs to the VAE to determine how the generated images are distributed through the encoded space and assess their potential for bolstering morphology prediction.

\section*{Acknowledgements}
This research was supported through funding from the Scheme for Promotion of Academic and Research Collaboration (SPARC) grant MHRD-18-0015.
MDW was supported by the University of Manchester and Rolls-Royce plc under the Engineering and Physical Sciences Research Council (EPSRC) grant EP/S022635/1 and Science Foundation Ireland (SFI) grant 18/EPSRC-CDT/3584.
CPR was supported by a University Research Fellowship of the Royal Society.
PJW was supported by the Henry Royce Institute for Advanced Materials, funded through EPSRC grants EP/R00661X/1, EP/S019367/1, EP/P025021/1 and EP/P025498/1.

\bibliographystyle{elsarticle-num}
\bibliography{references}

\end{document}